\documentclass[a4paper,twoside]{article} 
\usepackage{amsmath} 
\usepackage{amssymb} 
\usepackage{epsfig,float} 
\usepackage{pifont}
\usepackage{cite}

\input ijmpblat 
\newcommand{\nn}{\nonumber} 
\newcommand{\cuo}{CuO$_2$\ } 
\newcommand{\ie}{\textrm{i.e.}\ } 
\newcommand{\etal}{\textit{et al.}} 
\newcommand{\eg}{\textrm{e.g.}\ } 
\newcommand{\tc}{T_{\rm c}} 
\newcommand{\kb}{k_{_{\rm B}}} 
\newcommand{\eps}{\varepsilon} 
\newcommand{\egi}{\epsilon_{_{\rm Gi}}} 
\newcommand{\ecut}{\varepsilon_{_{\mbox{\scriptsize\ding{36}}}}} 
\newcommand{\Mathematica}{%
 \textit{Mathematica}$^{\mbox{\scriptsize\textcircled{\tiny{R}}}}$%
} 
 
\newcommand{\EM}{\mathsf{\hat\Sigma_{EM}}} 
\newcommand{\Reg}[1]{{\sf\widehat{Reg}_{#1}}} 
\newcommand{\Lop}{{\sf\hat{\hspace{3pt}L}}} 
\newcommand{\Cop}{{\sf\hat C}} 
\newcommand{\jour}[4]{{\nineit #1}\ {\ninebf #2},\ #3\ (#4)} 
\newcommand{\Fref}[1]{Fig.~\ref{#1}} 
\newcommand{\Eref}[1]{Eq.~(\ref{#1})} 
\newcommand{\Rref}[1]{Ref.~\citen{#1}} 
\newcommand{\PRL}{Phys. Rev. Lett.} 
\newcommand{\PR}{Phys. Rev.}

\begin{document} 
\runninghead{T. Mishonov \& E. Penev}%
            {Thermodynamics of Gaussian fluctuations and paraconductivity 
             in layered superconductors} 
%
\thispagestyle{empty} 
\setcounter{page}{1} 
\copyrightheading{Vol. 14, No. 32 (2000) 3831--3879} 
\vspace*{0.88truein} 
\fpage{1} 
\centerline{\bf THERMODYNAMICS OF GAUSSIAN FLUCTUATIONS AND} 
\vspace*{0.035truein} 
\centerline{\bf PARACONDUCTIVITY IN LAYERED SUPERCONDUCTORS} 
\vspace*{0.37truein}
\centerline{\footnotesize TODOR MISHONOV$^{\dag,\ddag,}$\footnote{ 
   Corresponding author: phone: (++32) 16 327 183, fax: (++32) 16 327 983,\\ 
   e-mail:~\texttt{todor.mishonov@fys.kuleuven.ac.be}; 
             \texttt{mishonov@rose.phys.uni-sofia.bg}
  }\ \ 
  {\normalsize and} EVGENI PENEV$^{\ddag}$ 
} 
\vspace*{0.015truein} 
\centerline{$^{\dag}$\footnotesize\it 
   Laboratorium voor Vaste-Stoffysica en Magnetisme, %
   Katholieke Universiteit Leuven 
} 
\baselineskip=10pt 
\centerline{\footnotesize\it 
   Celestijnenlaan 200 D, B-3001 Leuven, Belgium 
} 
\vspace*{0.015truein} 
\centerline{$^\ddag$\footnotesize\it 
   Department of Theoretical Physics, Faculty of Physics,%
   University of Sofia 
} 
\baselineskip=10pt 
\centerline{\footnotesize\it 
    5 J. Bourchier Blvd., 1164 Sofia, Bulgaria 
} 
\vspace*{0.225truein} 
\pub{16 July 2000}
 
\vspace*{0.21truein} 
\abstracts{ 
A detailed theoretical analysis of the Gaussian fluctuations of 
the order parameter in layered superconductors is performed within 
the Ginzburg-Landau (GL) theory. The available results for the 
Gaussian fluctuations are systematized and a number of novel formulae 
for the fluctuation magnetization, nonlinear magnetic 
susceptibility, heat capacity and high-frequency conductivity in 
layered superconductors are derived. 
We propose several new prescriptions: 
how to determine the life-time constant of fluctuation 
Cooper pairs $\tau_0,$ the in-plane coherence length $\xi_{ab}(0),$ 
the energy cutoff parameter $\ecut=c \hbar^2/2m_{ab}\xi^2_{ab}(0),$ 
and the Ginzburg number $\egi.$ 
%
It is demonstrated, for example, how the spectroscopy of the life-time 
can be used to verify the existence of depairing mechanisms in layered 
cuprates. The  ultraviolet regularization of the GL free energy 
is carried out by means of the well-known from the field theory 
$\zeta$-function method. We further show that the archetype of the 
latter method has its origin in the century of enlightenment and the 
novel result is that the fluctuation part of the thermodynamic variables 
of the layered superconductors can be expressed in terms of the Euler 
$\Gamma$-function and its derivatives. 
%
Universal scaling curves for the magnetic field dependence of 
the paraconductivity $\sigma_{ab}(\tc,B),$ fluctuation magnetization 
$M(\tc,B)$ and the heat capacity $C(\tc,B)$ are found for 
the quasi-2D superconductors at $\tc$ and further related to the 
form-factor of the Cooper pairs. 
}{}{} 
 
\vspace*{0.21truein} 
 
\vspace*{1pt}\textlineskip 
\section{Introduction} 
\vspace*{-0.5pt} 
\noindent 
The writing of this review is provoked by the progress in the 
study of fluctuation phenomena in high-$\tc$
superconductors.\cite{Ausloos} The small coherence lengths in the 
layered cuprates $\xi_{c} \ll \xi_{a} \simeq \xi_{b}$ give rise to 
a very high density of fluctuation degrees of freedom $\propto 
1/(\xi_{ab}^2(0)\xi_c(0))$ which makes the fluctuation effects 
easier to be observe in the high-$\tc$- rather than in the conventional 
superconductors. 
An intriguing feature of the fluctuation effects to be pointed out is 
that they can be observed even in the case when the interaction between 
fluctuations is vanishing or can be treated in a self-consistent 
manner. In such a case, for high quality crystals, the fluctuations 
are of Gaussian nature and their theory is very simple. A number of 
good experimental studies have already been performed in 
the Gaussian regime thus initiating the Gaussian fluctuation 
spectroscopy for high-$\tc$ materials. 
 
By spectroscopy here we imply only those experiments with trivial 
theory where every measurement provides an immediate information 
for some parameter(s) important for the material science or fundamental 
physics of these interesting materials. Half a century ago Landau used 
to speak about himself as being the greatest trivializator in the 
theoretical physics. At present, the Ginzburg-Landau (GL) theory 
(called by Ginzburg also $\Psi$-theory) is the adequate tool to 
describe the fluctuation phenomena in the superconductors. The 
parameters of the $\Psi$-theory, such as coherence lengths, 
relaxation time $\tau_{0,\Psi}$ of the GL order parameter $\Psi,$ 
the GL parameter $\kappa_{_{\rm GL}}=\lambda_{ab}(0)/\xi_{ab}(0)$ are 
also "meeting point" between the theory and the experiment. 
 
From one hand, these parameters are necessary for the description of the 
experimental data and from another hand they can be derived form the 
microscopic theory using the methods of the statistical mechanics. 
That is why the determination of the GL parameters is an important 
part of the investigations of every superconductor and the 
Gaussian fluctuation spectroscopy is an indispensable tool in 
these comprehensive investigations. 
 
The purpose of this review is to systematize the known classical results 
for the GL Gaussian fluctuations, to derive new ones when needed, and to 
finally give suitable for coding formulae necessary for the 
further development of the Gaussian spectroscopy. The derivation 
of all results is described in detail and trivialized to the level 
of the Landau-Lifshitz encyclopedia on theoretical physics,\cite{Landau9} 
the textbooks by Abrikosov\cite{Abrikosov} and Tinkham\cite{Tinkham} or 
the well-known reviews by Cyrot\cite{Cyrot} on the GL theory, 
by Bulaevskii\cite{Bulaevskii} 
concerning the layered superconductors with Josephson coupling, 
 and by Skocpol and Tinkham\cite{skoc} on the fluctuation phenomena in 
superconductors. 
The present work is intended as a review on the theoretical results 
which can be used by the Gaussian spectroscopy of fluctuations 
but no historical survey of the experimental research is attempted. 
Therefore we do explicitly refer to only a limited number of experimental 
studies in this field. 
Instead, the reader is referred to the citations-reach 
conference proceedings,\cite{Ausloos} but even therein a number of 
good works are probably not included. We do not cite directly 
even the epoch-creating paper by Bednorz and M\"uller but its spirit 
can be traced to every contemporary paper on high-$\tc$ 
superconductivity. Even to focus on the theoretical results 
related to fluctuation phenomena is a very difficult problem by itself 
and therefore, when referring to any result one should imply 
"to the best of our knowledge...". 
One of our goals was also to fill the gap between the textbooks and 
experimentalists' needs for a compilation of theoretical formulae 
written in common notations, appropriate for direct use. 
 
Of course, there is a great number of interesting physical situations 
especially related to vortices where the fluctuations are definitely 
non-Gaussian. Those problems fall beyond the scope of the present review and 
we include only some references from this broad field in the physics of 
superconductivity.\cite{NonGaussian} 
 
The review is organized as follows: in 
Sec.~2\ref{weakmagneticfields} the case of weak magnetic fields is 
considered and the thermodynamic variables are expanded in power 
series in the dimensionless magnetic field $h=B_z/B_{c2}(0).$ 
The standard notations for the thermodynamic variables in a layered 
superconductor are then introduced in Subsec. 2.1\ref{notations}, 
and Subsec.~2.2\ref{eulermaclaurinsummation} is dedicated to the 
Euler-MacLaurin summation formula in the form appropriate for the 
analysis of the GL results for the free energy and its ultraviolet (UV) 
regularization.
A systematic procedure to derive the results for a layered
superconductor from the results for a two-dimensional (2D)
superconductor is developed in Subsec.~2.3 and the action of the
introduced "layering" operator $\Lop$ is illustrated on the example of
the formulae for the paraconductivity. Further we consider the
static paraconductivity in case of perpendicular magnetic field as
well as the high frequency conductivity in zero magnetic field. The
power series for the nonlinear magnetic susceptibility and the
magnetic moment in the Lawrence-Doniach (LD) model are derived in
Subsec.~2.4 and the $\eps$-method for summation of such
divergent series is described in Subsec.~2.5. For practical purposes a
simple \textsc{fortran90} program is given in the Appendix. 

Further in Subsec.~2.6 we present the power series for the
differential susceptibility and general weak-magnetic-field expansion
formulae for the magnetization. Section~3 is dedicated to the study of
the strong magnetic fields limit. Firstly, in Subsec.~3.1 the general
formula for the Gibbs free energy in perpendicular to the layers
magnetic field is analyzed. The fluctuation part of the thermodynamic
variables is found then by differentiation in Subsec~3.2. Subsec.~3.3
is devoted to the self-consistent mean-field treatment of the
fluctuation interactions in the LD model. The important limit case of
an anisotropic 3D GL model is considered in Subsec.~3.4 where we
derive the Gibbs free energy and the fluctuation magnetic moment. In
Sec.~4 an account is given of the fitting procedure for the GL
parameters which rests on theoretical results and some recommendations
for the most appropriate formulae are also given for determination of
the cutoff energy $\ecut$ in Subsec.~4.1, the in-plane coherence
length $\xi_{ab}(0)$ in Subsec.~4.2, the Cooper pair life-time constant
$\tau_0$ in Subsec~4.3, and the 2D Ginzburg number in Subsec.~4.4.  All
new results derived throughout this review are summarized in Sec.~5
and some perspectives for the Gaussian spectroscopy are discussed as
well.
 
\section{Weak magnetic fields} 
\label{weakmagneticfields} 
 
\subsection{Formalism} 
\label{notations} 
\noindent 
Before embarking on a detailed analysis we shall briefly introduce all
entities entering the basic for our further considerations
quantity---the GL functional $G$ for the Gibbs free energy in external
magnetic field ${\bf H}^{\rm (ext)}$. For compliance with the previous
works we follow the standard notations in which $G$ reads
\begin{align}
  G[\Psi_{j,n} &(x,y), {\bf A} ({\bf r})] \nn\\ 
   =& \sum_{n=-\infty}^{+\infty} \sum_{j=1}^{N} \int dx\,dy 
     \Biggl\{ 
       \sum_{l=x,y} \frac{1}{2m_{ab}} 
       \left| 
         \left( 
               \frac{\hbar}{i} \frac{\partial}{\partial x_l} - e^* A_l 
         \right) \Psi_{jn} 
       \right|^2 \nn \\
 &   +  a_0 \epsilon \left|\Psi_{j,n}\right|^2 
     + \frac{1}{2} \tilde b \left|\Psi_{j,n}\right|^4  
     +  a_0 \gamma_j \left| \Psi_{j+1,n} - \Psi_{j,n} 
        \exp 
            \left(\frac{ie^*}{\hbar} 
         \int_{z_{j,n}}^{z_{j+1,n}} 
         A_z dz \right) \right|^2 \Biggr\} \nn \\
 & +  \int \frac{1}{2 \mu_0} 
 \left( \nabla \times {\bf A} - \mu_0  {\bf H}^{\rm (ext)}\right)^2 
 dx\,dy\,dz,
\label{GLGibbs}  
\end{align} 
with \textbf{A} being the vector potential of the magnetic field 
$B=\nabla\times \textbf{A}.$ 
 
\begin{figure}[t] 
  \begin{center} 
    \epsfig{file=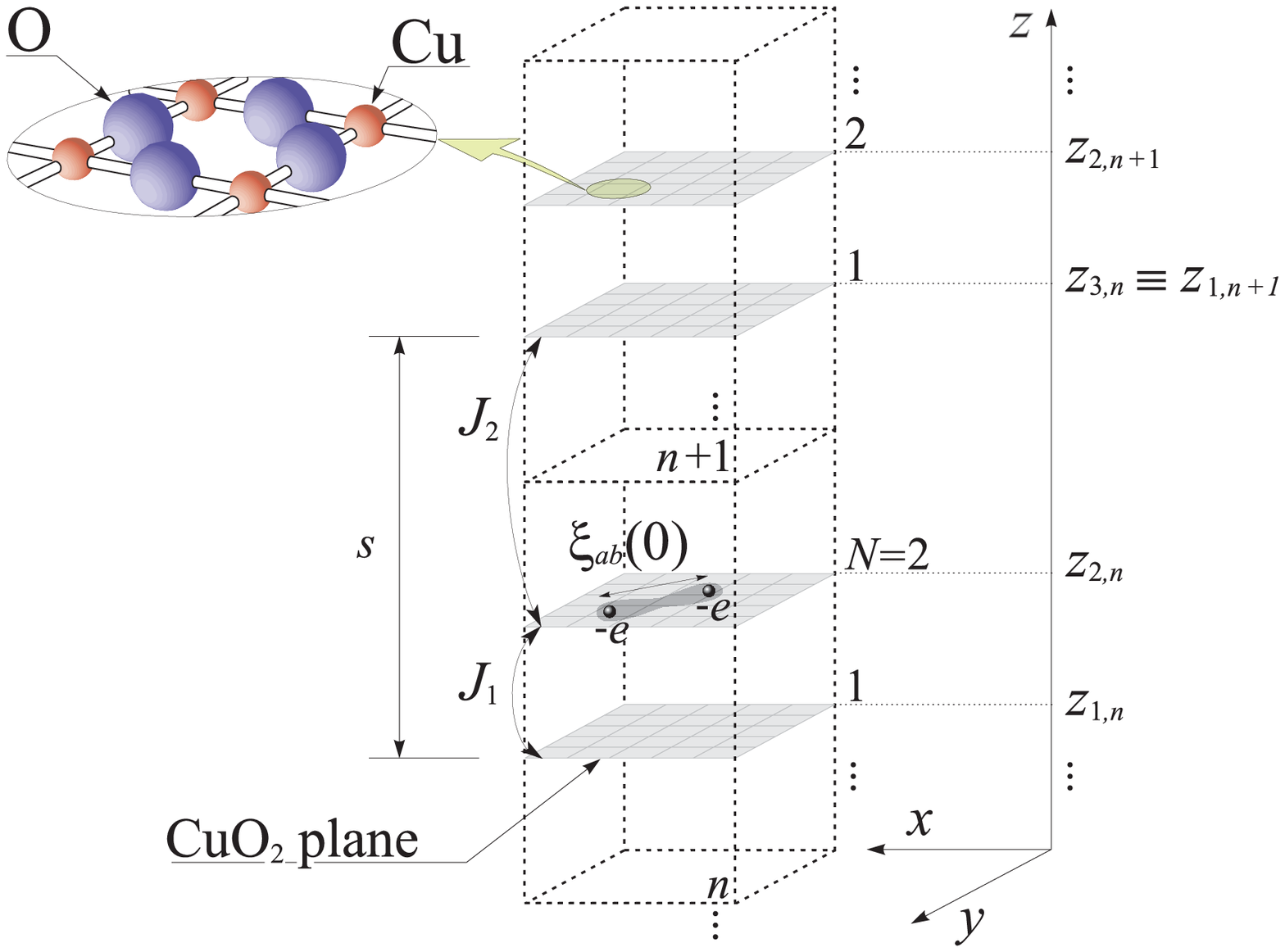,width=10cm} 
  \end{center} 
  \vspace*{13pt} 
  \fcaption{%
    Pictorial representation of material parameters that enter 
    \protect{\Eref{GLGibbs}. The $z$-coordinates of the \cuo planes, 
    satisfying periodic boundary conditions, are $z_{N+1,n}=z_{1,n+1}$}, where 
    $n$ labels the unit cell  and $j=1,\ldots, N$ is the index of the \cuo 
    plane within each unit cell. Periodicity in $c$-direction is designated by 
    $s,$ \ie $0 \le z_{j,0} \le s.$ $\xi_{ab}(0)$ stands for the in-plane 
    coherence length extrapolated to $T=0.$ The Josephson coupling energies 
    between the neighboring planes $J_j = a_0\gamma_j$ are parameterized via 
    dimensionless quantities $\gamma_j.$ Lastly, the effective charge and the 
    in-plane effective mass of the Cooper pairs are, respectively, 
    $|e^*|=2|e|$ and $m_{ab}.$ 
    \label{fig:1} } 
\end{figure}

The material parameters in this sizable expression are illustrated in 
\Fref{fig:1}, thus we only need to note that the GL potential 
$a(\epsilon)=a_0 \epsilon $ is parameterized by 
$a_0=\hbar^2/2m_{ab}\xi_{ab}^2(0)$, and 
$\epsilon \equiv \ln(T/\tc)\approx (T-\tc)/\tc$ is the reduced 
(dimensionless) temperature. If not otherwise stated we shall make use 
of the SI units, thus the magnetic permeability of vacuum 
$\mu_0=4\pi\times 10^{-7}.$ 
 
Here we will restrict ourselves to the study of fluctuations in the Gaussian 
regime in the normal phase not too close to the critical line $H_{c2}(T).$ 
In this case the nonlinear term in $G[\Psi,\textbf{A}]$ is negligible, 
$\tilde b \vert\Psi\vert^4/2 \rightarrow 0.$ 
For the normal phase the magnetization is also very small and with high accuracy 
$\mu_0 {\bf H^{\rm (ext)}}\approx{\bf B}= \mu_0({\bf H} + 
{\bf M} ) \approx \mu_0 {\bf H}.$ 
 
To begin with, consider the simplest case of zero external magnetic field 
${\bf H}^{\rm (ext)} = 0.$ 
Given the above assumption for $\Psi,$ the GL functional is a quadratic form and one 
needs to sum over all eigenvalues of the energy spectrum 
\begin{equation} 
  \eps_j ({\bf p},p_z) = \frac{{\bf p}^2}{2m_{ab}} + \eps_{cj}(p_z), \quad 
  \eps_{cj}(p_z)       = a_0 \omega_j^{(N)} (\theta), 
\label{spectrum} 
\end{equation} 
where $\varepsilon_{cj}(p_z)$ are the tight-binding energy bands 
describing the motion of Cooper pair in $z$ ($c$) direction, ${\bf 
p}= (p_x,p_y)$ is the in-plane ($ab$-plane) momentum of the 
fluctuating Cooper pairs and $\theta=p_zs/\hbar \in (0, 2\pi)$ is 
the Josephson phase. For a single layered material, $N=1$, 
this corresponds to the well known Lawrence-Doniach model,\cite{LD} 
\begin{equation} 
  \omega_1^{\rm (LD)}(\theta)=2\gamma_1(1-\cos \theta) 
\end{equation} 
while the case $N=2$ is the Maki-Thompson (MT) model,\cite{Maki} 
proposed independently  by Hikami and Larkin\cite{hikami-larkin} as well, 
\begin{equation} 
  \omega_j^{\rm (MT)}(\theta)=\gamma_1 + \gamma_2 
  + (-1)^j \sqrt{ \gamma_1^2+\gamma_2^2+2\gamma_1\gamma_2\cos \theta }, 
  \qquad j=1,2. 
\label{bilayer} 
\end{equation} 
Thus, the sum over the energy spectrum gives 
\begin{equation} 
  G[\Psi]= \sum_{{\bf p},p_z,j} 
  \left(\varepsilon_j({\bf p},p_z)+a\right)\, 
  \left|\Psi_{{\bf p},p_z,j}\right|^2, 
\label{Gmomentum} 
\end{equation} 
where $ \Psi_{{\bf p},p_z,j}$ is the  wave function of the superconducting 
condensate in momentum space. We use standard periodic boundary conditions 
for a bulk domain of volume $V=L_x \times L_y \times L_z$ which give 
\begin{equation} 
  \sum_{{\bf p}}= L_x L_y \int \frac{dp_x\,dp_y}{(2\pi\hbar)^2}, 
    \qquad 
  \sum_{p_z}= L_z \oint \frac{d\theta}{2\pi s}. 
\label{psum} 
\end{equation} 
In order to calculate the fluctuation part of the Gibbs free 
energy $G(T)$ at zero magnetic field, one usually solves 
for every point in the momentum space ${\bf p},p_z,j$ the Gaussian integral 
\begin{align}
  \exp\left[-\frac{G}{\kb T}\right] & =  
  \iint \frac{d\Psi^{\prime}\;d\Psi^{\prime \prime}}{2\pi} 
   \exp 
    \left\{ - \frac{\varepsilon + a}{\kb T} 
      \left[\left( \Psi' \right)^2 + \left( \Psi'' \right)^2 \right] 
    \right\} \nn \\ 
 & =  \frac{1}{2} \int_0^{\infty} \exp \left[ - \frac{\eps + a}{\kb T} 
       \rho \right] d\rho 
     = \frac{\kb T/2}{\eps + a}, 
\end{align}
where $\Psi = \sqrt{\rho}\,e^{i\varphi}\equiv\Psi' + i\Psi'',$ 
and $\varphi \in (0,2 \pi).$ 
Making use of this auxiliary result the calculation of the fluctuation 
part of the Gibbs free energy reduces to summation over the spectrum 
of an effective Hamiltonian, \ie 
\begin{equation} 
  G = -\kb T \sum_{{\bf p},p_z,j} \ln
  \frac{\kb T/2}{\eps_j({\bf p},p_z) + a}, 
\label{DeltaG} 
\end{equation} 
or, taking into account Eqs.~(\ref{spectrum}) and (\ref{psum}), 
\begin{equation} 
  \frac{G}{V} = \kb T \int\frac{ 
  d\left( \pi {\bf p}^2 \right)}{(2 \pi \hbar)^2}\, {1\over N} 
  \sum_{j=1}^N \oint \frac{d\theta}{2 \pi s} \ln \left[ 
  \frac{({\bf p}^2/2m_{ab})+a_0 \omega_j^{(N)} (\theta) + a_0\epsilon}{a_0} 
  \frac{a_0}{\frac{1}{2}\kb T} \right]. 
\label{Gfl} 
\end{equation} 
In view of the further calculations it is also useful to introduce a 
dimensionless in-plane kinetic energy 
\begin{equation} 
  {\tilde x} = \frac{{\bf p}^2}{2m_{ab}a_0} 
             = \left(\frac{\xi_{ab}(0){\bf p}}{\hbar}\right)^2 \in (0,c), 
\end{equation} 
bound by a dimensionless cutoff parameter $c$ which we consider to be 
an important parameter of the GL theory when applied to 
copper oxide superconductors. Later in Sec~4.1 we demonstrate how the value 
of the dimensional cutoff energy $\ecut$, 
\begin{equation} 
  \frac{{\bf p}^2}{2m_{ab}} < \ecut = c a_0 =\frac{p_c^2}{2m_{ab}}, 
\end{equation} 
can be determined by fitting to the experimental data. 
An immediate simplification to \Eref{Gfl} can be achieved by dropping the 
${1\over 2} \kb T/a_0 \rightarrow {\rm const}$ multiplier in the argument of 
the logarithm as it is irrelevant for the critical behavior of the material. 
Furthermore, since fluctuational observables are related to non analytical 
dependence of the Gibbs free energy on the reduced temperature, we 
can substitute $T=\tc(1+\epsilon) \approx \tc$ and the free energy per 
unit volume $F(\epsilon)$ is cast in more elegant form, 
\begin{equation} 
  F(\epsilon) \approx \frac{G}{L_x L_y L_z}= 
  \frac{\kb \tc}{4 \pi \xi^2_{ab}(0)} 
  \frac{N}{s} \int_{0}^{c} d{\tilde x} 
  \frac{1}{N} \sum_{j=1}^{N}\oint \frac{d\theta}{2\pi} 
     \ln \left( 
       {\tilde x} + \omega_j^{(N)}(\theta) + \epsilon 
     \right), 
\label{Fe} 
\end{equation} 
that could easily include the $(1+\epsilon)$-factor in all cases when 
necessity appears. The physical meaning of this 
important for our further considerations expression is fairly transparent: 
one has to integrate with respect to the  Josephson phase $\theta,$ which 
describes the motion of Cooper pairs in $c$-direction, and to take 
into account as many different Cooper pair energy bands as are there the 
different superconducting layers per unit cell. 
Finally, integration with respect to the in-plane Cooper pair kinetic energy 
is to be carried out. 
 
Consider now the important case of an external magnetic field applied 
parallel to the $c$-direction, \ie perpendicular to the \cuo planes, 
${\bf B} = (0,0,B).$ In this case, 
the in-plane kinetic energy of the Cooper pairs acquires oscillator 
spectrum,\cite{Landau3} corresponding to the quantum mechanical problem of 
an electron in an external magnetic field,\cite{Landau9} 
\begin{equation} 
  {{\bf p}^2\over 2m_{ab}}\rightarrow 
  \hbar \omega_c \left(n + {1\over2}\right), 
\end{equation} 
where $n=0,1,2,3,\dots$\ is a non-negative integer and  $\omega_c=\left| e^* 
\right| B/m_{ab}$ is the cyclotron frequency. The integration over the 
momentum space is thus reduced to summation over oscillator energy levels 
\begin{equation} 
  \int_{\left|{\bf p}\right|<p_{c}} 
  \frac{d^2 {\bf p}}{(2\pi\hbar)^2} \rightarrow \frac{B}{\Phi_0} 
  \sum_{n=0}^{n_c-1}, 
\end{equation} 
where $n_c\equiv c/2h$ and $\Phi_0=2\pi\hbar/\left| e^* \right|= 
2.07$~${\rm fT\,m^2}$ is the flux quantum. The energy cutoff is to be 
applied now to the oscillator levels,\cite{Landau9} 
$\hbar\omega_c (n_c + {1\over 2})=ca_0.$ 
Let us recall that the equation for the upper critical field 
$H_{c2}(T)$ within the GL theory is nothing but the equation for annulment 
of the lowest energy level, ${1\over 2} \hbar \omega_c + a(\epsilon)=0.$ 
Thereby introducing the upper critical field linearly extrapolated to 
zero temperature, 
\begin{equation} 
  \mu_0 H_{c2}(0) = B_{c2}(0) \equiv - \tc 
  \left.\frac{dB_{c2}(T)}{dT}\right|_{\tc} = 
  {\Phi_0\over 2\pi\xi_{ab}^2(0)}, 
\end{equation} 
and the dimensionless reduced magnetic field, 
\begin{equation} 
  h\equiv\frac{B}{B_{c2}(0)} = \frac{H}{H_{c2}(0)}, 
\end{equation} 
we obtain a linear approximation for the critical line about  $\tc,$ 
\begin{equation} 
  h_{c2}(\epsilon) = {H_{c2}(T) \over H_{c2}(0)} \approx -\epsilon \ll 1 . 
\end{equation} 
With the help of the dimensionless variables introduced so far it is easily worked 
out that the influence of the external magnetic field is reduced to discretization 
of the dimensionless in-plane kinetic energy, 
\begin{equation} 
  {\tilde x} \rightarrow h(2n+1) 
\end{equation} 
and the integrals of an arbitrary function $f$ with respect to $\tilde x$ 
are converted to sums, 
\begin{equation} 
  \int_{0}^{c} f({\tilde x}) d{\tilde x} 
  \rightarrow 2h \sum_{n=0}^{n_c-1} f(h(2n+1)). 
\end{equation} 
In fact, Max Planck discovered the quantum statistics of the black-body radiation 
using the same replacement. Applying this procedure to the previously derived free 
energy at zero magnetic field, \Eref{Fe}, we obtain 
\begin{equation} 
  F(\epsilon)\rightarrow F(\epsilon,h) =\frac{\Delta G}{L_x L_y L_z}, 
\end{equation} 
\begin{equation} 
  F(\epsilon,h)= \frac{\kb \tc}{4 \pi \xi_{ab}(0)^2} 
  \frac{N}{s} 2h 
  \sum_{n=0}^{n_c-1} \frac{1}{N} \sum_{j=1}^{N}\oint \frac{d\theta}{2\pi} 
  \ln \left[ h(2n+1) + \omega_j^{(N)}(\theta) + \epsilon \right]. 
\label{Feh} 
\end{equation} 
This expression represents the starting point for all further considerations. 
As a first step we address in the next section the Euler-MacLaurin method and its 
application to the sum over the Landau levels which appears in \Eref{Feh}. 
 
\subsection{Euler-MacLaurin summation for the free energy} 
 
\label{eulermaclaurinsummation} 
\noindent 
Near the critical temperature, when $\epsilon\ll c,$ one can consider formally 
$c\rightarrow \infty$, and $n_c(h)\approx c/2h\rightarrow \infty.$ 
Within such a local approximation the previous finite sums are transformed 
into infinite ones, 
\begin{equation} 
  2h \sum_{n=0}^{\infty} f(\epsilon+h(2n+1)) = 
  \EM \int\limits_{\epsilon}^{\infty} f({\tilde x}) d{\tilde x}, 
\label{sum-int} 
\end{equation} 
where 
\begin{align} 
  \EM \equiv \frac{ h \frac{\partial}{\partial \epsilon}}%
                   {\sinh\!\left( h \frac{\partial}{\partial\epsilon}\right)} 
  & =  \sum_{n=0}^{\infty} (-1)^n 
        \frac{2}{\pi^{2n}} \left( 1- \frac{1}{2^{2n-1}} \right) \zeta(2n) 
        \left(h \frac{\partial}{\partial\epsilon}\right)^{2n} \nn \\ 
  & =  1 - {1  \over     6} h^2 \frac{\partial^2}{\partial \epsilon^2} 
         + {7  \over   360} h^4 \frac{\partial^4}{\partial \epsilon^4} 
         - {31 \over 15120} h^6 \frac{\partial^6}{\partial \epsilon^6} 
         + \cdots 
\label{E-M} 
\end{align} 
is the Euler-MacLaurin operator for summation of series, in which we employ 
the Riemann and Hurwitz zeta functions, respectively, 
\begin{equation} 
  \zeta(\nu)= 1 + {1\over 2^{\nu}} + {1\over 3^{\nu}} + \cdots =\zeta(\nu,1), 
  \qquad 
  \zeta\left(\nu,z\right) =\sum_{n=0}^{\infty} \,\frac{1}{\left(n+z\right)^{\nu}}, 
\end{equation} 
instead of Bernoulli numbers. The $\EM$ operator can be easily obtained exploiting 
the exponential representation of the standard translation operator $\sf\hat T$, 
whose action is defined as follows 
\begin{equation} 
  f(b+\epsilon) = \left. {\sf\hat T}_z(b)f(z) \right|_{z=\epsilon} 
                = \left. \exp\!\left(b\frac{\partial}{\partial z}\right)f(z) 
                         \right|_{z=\epsilon}. 
\end{equation} 
If summed up from zero to infinity the above expression would give an infinite 
geometric progression, 
\begin{equation} 
  \sum_{n=0}^{\infty} \left[ {\sf\hat T}_z(b) \right]^n 
  = \sum_{n=0}^{\infty} \left[ \exp\!\left(b\frac{\partial}{\partial z}\right)
    \right]^n 
  = {1 \over 1 - \exp\!\left(b {\partial \over \partial z}\right)}. 
\label{progression} 
\end{equation} 
 
Let us introduce now the fluctuational part of the heat capacity, 
\begin{equation} 
 C(\epsilon) = - {1\over \tc} 
   \frac{\partial^2}{\partial \epsilon^2} F(\epsilon). 
\end{equation} 
Using this physical observable and \Eref{Feh}, one can extract the magnetic 
field dependent part of the free energy, 
\begin{align} 
F(\epsilon,h)-F(\epsilon) 
 & =  \sum_{n=1}^{\infty} (-1)^{n-1} \frac{2}{\pi^{2n}} 
      \left( 
        1- \frac{1}{2^{2n-1}} 
      \right) 
      \zeta(2n) h^{2n} \frac{\partial^{2(n-1)}}{\partial\epsilon^{2(n-1)}} 
      \tc C (\epsilon)  \nn \\ 
 & =  \left[ 
        {1\over 6} h^2 
        - {7\over 360} h^4\frac{\partial^{2}} {\partial\epsilon^{2}} 
        + {31\over 15120}h^6\frac{\partial^{4}} {\partial\epsilon^{4}} 
        - \cdots 
      \right] \tc C (\epsilon). 
\label{Fmagn} 
\end{align} 
It is then straightforward to calculate the magnetization $M$ and the 
nonlinear susceptibility defined as $\chi(\epsilon,h)\equiv M/H,$ \ie 
\begin{equation} 
  M =-\frac{\partial [F(\epsilon,h) - F(\epsilon)]} {\partial B} 
    = \chi(\epsilon,h) H. 
\label{M} 
\end{equation} 
For the Meissner-Ochsenfeld (MO) state, for example, $\chi^{\rm (MO)} = -1.$ 
Importantly, \Eref{M} incorporates the \textit{regularized} free energy which, 
by virtue of \Eref{Fmagn}, does not contain zero magnetic field part, 
\begin{equation} 
  F_{\rm reg}(\epsilon,h) \equiv F(\epsilon,h) -F(\epsilon) 
                          := \Reg{EM} F(\epsilon) 
\end{equation} 
where 
\begin{equation} 
 \Reg{EM} = \EM - {\sf\hat{\sf 1}} = 2h \sum_{n=0}^{n_{c}} 
            - \int_{0}^{c} d{\tilde x} 
\end{equation} 
is the Euler-MacLaurin regularization operator. This method was 
applied\cite{schmidt,schmid,vschmidt} for calculation of zero-field limit of 
the magnetic susceptibility, cf. \Rref{Landau9}. 
Hence, inserting the $h^2$-term of \Eref{Fmagn} into \Eref{M} 
one finds\cite{manolo} for $H \rightarrow 0$ 
\begin{equation} 
  \chi(\epsilon)= - \frac{\mu_0 \tc C(\epsilon)}{3B_{c2}^2(0)} 
   = -{4 \pi^2 \mu_0 \over 3\Phi_0^2}\xi_{ab}^4(0) \tc C (\epsilon). 
\label{chiC} 
\end{equation} 
For illustration, let us analyze how this relation between susceptibility 
and heat capacity can be applied to the LD model. 
For arbitrary multilayered structure we can calculate the curvature of the 
lowest dimensionless energy band in $c$-direction 
$\omega_1(\theta)=\eps_{c1}(p_z)/a_0,$ 
\begin{equation} 
  r \equiv 2 
  \frac{\partial^2}{\partial \theta^2} \omega_1(\theta)|_{\theta=0}. 
\end{equation} 
According to this definition $r$ parameterizes the effective mass in $c$-direction 
$m_c$ for an anisotropic GL model, $|p_z| \ll \pi \hbar/s$, 
\begin{equation} 
  \varepsilon_{c1}\approx a_0 \frac{r}{4} \theta^2 = \frac{p^2_z}{2m_c}. 
\label{parabolic} 
\end{equation} 
It is now easily realized that the identity holds true, 
\begin{equation} 
  r = \left(\frac{2 N \xi_c(0)}{s}\right)^2, 
\label{LDparameter} 
\end{equation} 
which for $N=1$ is the LD parameter $r$ that 
determines the effective dimensionality of the superconductor, cf. 
the review by Varlamov \etal\cite{varlall} In many other 
studies, \eg \Rref{skoc}, the wave vector ${\bf k}={\bf p}/\hbar$ 
has been used as well. 
In terms of the latter, for the dimensionless kinetic energy in the 
long-wavelength approximation we have 
\begin{equation} 
  \frac{\left[\varepsilon_{ab}(\hbar {\bf k}) 
  + \varepsilon_{c1}(\hbar k_z) \right]}{a_0} 
    \approx\xi_{ab}^2(0) {\bf k}^2 + \xi_c^2(0) k_z^2 
  = \xi_{ab}^2(0) {\bf k}^2 + r \theta^2, 
\end{equation} 
where $\eps_{ab}(\hbar {\bf k})$ represents the in-plane part of the kinetic energy. 
Let us mention that the LD model is not only applicable to single 
layered cuprates with $N=1,$ but is it also to bi-layered cuprates ($N=2$) in 
the limit cases $\gamma_1 \simeq \gamma_2$ as well as in the 
case $\gamma_1 \ll \gamma_2$ when formally $N=1$. That is why we 
use in our formulae an effective periodicity of the LD-model 
\begin{equation} 
  s_{\rm eff} = \frac{s}{N}. 
\label{seff} 
\end{equation} 
For completeness we list below without deriving some of the well-known results 
 within the  LD-model. The single energy band has the form 
\begin{align} 
  \varepsilon_c(p_z) & = a_0 \omega_1 \left(\frac{p_z s}{2\pi \hbar}\right) 
                       = \frac{\hbar^2}{m_c (s/N)^2} 
                         \left(1-\cos \theta\right), \mbox{ where}\\ 
  \omega_1(\theta)   & = {1\over 2} r \left(1-\cos \theta\right) = 
                         r \sin^2{\theta\over2}, 
\label{omega1} 
\end{align} 
being parameterized by the Josephson coupling energy 
\begin{equation} 
  J_1 = a_0 \gamma_1=\frac{\hbar^2}{m_c (s/N)^2}, \qquad 
    r = 4\gamma_1 
      = \left( 2 N \xi_c(0)/s \right)^2. 
\end{equation} 
For the heat capacity one has 
\begin{equation} 
  C^{\rm (LD)} (\epsilon) = {\kb\over 4\pi \xi^2_{ab}(0)} {N \over s} 
  {1\over \sqrt{\epsilon \;(\epsilon + r)}}, 
\label{CLD} 
\end{equation} 
and the magnetic susceptibility for a weak magnetic field applied 
in $c$-direction, according to Tsuzuki\cite{tsuzuki} and Yamayi,\cite{yamayi}
reads as 
\begin{equation} 
  -\chi^{\rm (LD)}(\epsilon)= {\pi \over 3}\mu_0 
  \frac{\kb \tc}{\Phi_0^2}\xi_{ab}^2(0) {N\over s} \;{1\over 
  \sqrt{\epsilon}}\; {1\over\sqrt{\epsilon +r}} ={1\over 
  6}\frac{M_0}{H_{c2}(0)}{1\over \sqrt{\epsilon \;(\epsilon + r)}}, 
\label{chiLD} 
\end{equation} 
where 
  \begin{equation} M_0\equiv\frac{\kb \tc}{\Phi_0} \frac{N}{s}. 
\label{M0} 
\end{equation} 

Before proceeding we feel it appealing to make some 
technical remarks concerning the representation of the general formulae 
for fluctuations in arbitrary layered superconductor. To be specific, 
we shall demonstrate how the expressions for the magnetic 
susceptibility, \Eref{chiLD}, and heat capacity, \Eref{CLD}, within the 
LD model can be obtained as special cases of a general procedure 
described in the next subsection. 
 
\subsection{Layering operator $\Lop$ illustrated on the example 
            of paraconductivity} 
 
\noindent 
In the general formula for the density of the free energy, \Eref{Fe}, 
the energies related to motion in $c$-direction $\eps_{cj}(p_z)$ 
enter the final result solely via the fragment 
$\epsilon + \omega_{j}(\theta).$ Thus, in all such cases one can first solve 
the corresponding 2D problem and then for a layered superconductor the 
result can be derived by merely averaging the 2D result with respect to 
the motion of the fluctuation Cooper pairs in perpendicular to the layers 
direction. Formally, this method reduces to introducing a \textit{layering} 
operator $\Lop$ acting on functions of $\epsilon$; \eg for the conductivity 
one would have the relation 
\begin{equation} 
  \sigma(\epsilon)= \Lop \sigma^{\rm (2D)} (\epsilon) 
  \equiv 
  {1 \over N} \sum_{j=1}^N \oint \frac{d\theta}{2\pi} 
  \sigma^{\rm (2D)}\!\left(\epsilon+\omega_{j}^{(N)}(\theta) \right). 
\label{layer} 
\end{equation} 
In terms of the so introduced operator $\Lop$ the expression for the free 
energy, \Eref{Feh}, takes the form 
\begin{equation} 
F(\epsilon)= F_0 \int_{0}^{c} d{\tilde x} 
    \sum_{j=1}^{N} 
     \Lop\ln\!\left( {\tilde x} + \omega_j^{(N)}(\theta) + \epsilon \right), 
  \qquad 
  F_0 \equiv \frac{\kb \tc}{4 \pi \xi^2_{ab}(0)} \frac{N}{s}. 
\label{lFe} 
\end{equation} 
Besides for thermodynamic variables this operator works also for 
the fluctuation in-plane conductivity. Conforming with the work 
of Hikami and Larkin,\cite{hikami-larkin} for the conductivity within the 
LD model we have to integrate the 2D conductivity with respect 
to the Josephson phase, 
\begin{equation} 
  \sigma(\epsilon)= \Lop^{\rm (LD)} \sigma^{\rm (2D)} 
  (\epsilon) \equiv \oint \frac{d\theta}{2\pi} \sigma^{\rm 
  (2D)}\!\left(\epsilon+{1\over2}r(1-\cos \theta) \right). 
\label{layerLD} 
\end{equation} 
Given a system with independent 2D layers, having density in $c$-direction $N/s$, 
for zero magnetic field we have to average the 
well-known Aslamazov-Larkin expression for the static (zero-frequency) conductivity, 
\begin{equation} 
  \sigma_{\rm AL}(\epsilon)={e^2\over 16 \hbar}{N\over s}{1\over \epsilon} 
  = {\pi\over 8} R_{\rm QHE}^{-1}{N\over s}{1\over \epsilon}, 
\label{ALformula} 
\end{equation} 
where $R_{\rm QHE} \equiv 2\pi\hbar/e^2=25.813$~k$\Omega.$ 
A simple integration gives 
\begin{equation} 
  f_{\rm LD}(\epsilon;r) \equiv \Lop^{\rm (LD)} 
  {1\over\epsilon}= \oint \frac{d\theta } {2\pi} \; {1\over 
  \epsilon +{1\over 2} r\left(1 - \cos \theta \right)} = 
  \frac{1}{\sqrt{\epsilon (\epsilon + r)}}. 
\label{LD1} 
\end{equation} 
We note that this integral determines both the heat capacity and magnetic 
susceptibility for the LD model and is widely used for fitting 
to experimental data. Another important integral is 
\begin{align} 
    \Lop^{\rm (LD)}\int_{\epsilon}^{c} \ln \tilde \epsilon \; 
    d\tilde \epsilon 
 & =  \int_{\epsilon}^{c} 
    2\ln\!\left(\frac{\sqrt{\tilde\epsilon}+\sqrt{\tilde\epsilon+r}}{2}\right) 
    d \tilde \epsilon \nn \\ 
 & =  \left.\left[(2\tilde\epsilon + r) 
    \ln\!\left(\sqrt{\tilde\epsilon} + \sqrt{\tilde\epsilon+r}\right) 
    - \sqrt{\tilde\epsilon(\tilde\epsilon+r)} - 
     \ln(4)\;\tilde\epsilon\right]\right|_{\epsilon}^{c}, 
\label{LDxlnx} 
\end{align} 
which is used in representing the free energy at zero magnetic field, \Eref{Fe}, 
cf. also Eqs.~(\ref{theresult}) and (\ref{weakfieldM1}) below. 
Further, the $\epsilon$-derivative of this equation, 
\begin{equation} 
  \Lop^{\rm (LD)} \ln \epsilon= 
  \oint \frac{d\theta } {2\pi} 
  \ln \left( \epsilon +{1\over 2} r\left(1 - \cos \theta \right) \right)= 
  2\ln\!\left( 
  \frac{\sqrt{\epsilon} 
               +\sqrt{\epsilon+r}}{2} 
  \right), 
\label{LDln} 
\end{equation} 
is important for the calculation of the fluctuation part of the entropy and 
the density of fluctuation Cooper pairs. We provide also two other integrals 
employed in calculating the magnetoconductivity\cite{varlall} 
\begin{align} 
  \Lop^{\rm (LD)} {1\over\epsilon^2}  & =  
  -\Lop^{\rm (LD)}\frac{\partial}{\partial \epsilon} {1\over\epsilon} 
  = \frac{\epsilon+{1\over2}r}{[\epsilon(\epsilon+r)]^{3/2}}, \\ 
  \Lop^{\rm (LD)} {1\over\epsilon^3}  & =  
  {1\over 2} \frac{\partial^2}{\partial \epsilon^2}\Lop^{\rm (LD)} 
  {1\over\epsilon} =\frac{(\epsilon+r)\epsilon+{3\over8}r^2} 
  {[\epsilon(\epsilon+r)]^{5/2}}. 
\label{LD3} 
\end{align} 
Analogously, in the Maki-Thompson bi-layered model\cite{Maki} ($N=2$) 
one summation should precede the integration,\cite{manolo}
\begin{equation} 
  \sigma(\epsilon)= \Lop^{\rm (MT)} \sigma^{\rm (2D)} 
  (\epsilon) = \frac{1}{2} \oint \frac{d\theta}{2\pi} \left[ 
  \sigma^{\rm (2D)}\!\left(\epsilon + \omega_{1}(\theta)\right)+ 
  \sigma^{\rm (2D)}\!\left(\epsilon + \omega_{2}(\theta)\right) 
  \right], 
\label{layerMT} 
\end{equation} 
\ie in order to  calculate the conductivity\cite{Maki} and 
susceptibility\cite{manolo} we have to add the terms 
\begin{equation} 
  \frac{1}{\epsilon+\omega_1}+\frac{1}{\epsilon+\omega_2} 
  =\frac{2\epsilon+(\omega_1+\omega_2)} 
  {\epsilon^2+(\omega_1+\omega_2)\epsilon+\omega_1\omega_2}. 
\end{equation} 
In this expression both $\omega_1+\omega_2$ and $\omega_1\omega_2$ are rational, 
cf. \Eref{bilayer}, and the integral (\ref{layerMT}) 
is reduced to the integral (\ref{LD1}), 
\begin{align} 
  \Lop^{\rm (MT)}{1\over\epsilon} & = 
  \frac{\epsilon+\gamma_1+\gamma_2} {\sqrt{ \epsilon 
  \left[\epsilon+2(\gamma_1+\gamma_2)\right] 
  \left(\epsilon+2\gamma_1\right)\left(\epsilon+2\gamma_2\right)}} 
  \label{MT1} \nn \\ 
 & = \frac{\epsilon+{1\over 2}rw}{\sqrt{ \left(\epsilon^2+rw\epsilon\right) 
  \left(\epsilon^2+rw\epsilon+{1\over 4}r^2w\right)}} 
  \equiv f_{\rm MT}(\epsilon,h; r,w), 
\end{align} 
where 
$$ 
  r \equiv 4\frac{2}{{1\over\gamma_1}+{1\over\gamma_2}}, 
  \quad u\equiv\frac{J_{\rm max}}{J_{\rm min}} 
  = \frac{\gamma_{\rm max}}{\gamma_{\rm min}}, 
    \quad w\equiv{1\over 4} 
    \left(2+\frac{\gamma_{1}}{\gamma_{2}}+\frac{\gamma_{2}}{\gamma_{1}}\right) 
  = {1\over 4}\left(2+u+{1\over u}\right). 
$$ 
Such a form, involving the $w$ parameter, is convenient for fitting the 
experimental data since for both $w=1$ and $w\gg1$ cases the 
LD approximation holds true, which is often found to give a satisfactory 
explanation of the experimental observations. For more detailed discussion 
 the reader is referred to \Rref{Vidal}. 
The inverse relations for the above introduced parameters read as 
\begin{equation} 
  u = \left(2w-1\right)+2 \sqrt{w(w-1)}, \quad 
  \gamma_{\rm min}=\frac{1}{2}\left(1+{1\over u}\right){r\over 4},\quad 
\gamma_{\rm max}=\frac{1}{2}\left(1+{u}\right){r\over 4}, 
\end{equation} 
and 
\begin{equation} 
  J_{\rm max}=a_0\gamma_{\rm max}=\frac{\hbar^2}{2m_{ab}\xi^2_{ab}(0)} 
  \frac{1}{2}\left(1+u\right){r\over 4}. 
\end{equation} 
Importantly, at known effective mass $m_{ab}$ the last equation gives the 
possibility one to determine the Josephson coupling energy 
between double \cuo  planes. 
 
Let us now illustrate in more details the action of the $\Lop$ operator. 
Towards this end we consider the famous Aslamazov-Larkin\cite{AL1} formula for 
the 2D conductivity $\sigma_{ab}^{\rm (2D)}(\epsilon)$, the results for the 
susceptibility $-\chi_{ab}^{\rm(2D)}(\epsilon)$ due to 
V.~Schmidt,\cite{vschmidt} A.~Schmid\cite{schmid} and 
H.~Schmidt,\cite{schmidt} 
and Ferrell\cite{Ferrell} and Thouless\cite{Thouless} fluctuation part of the 
heat capacity $C^{\rm(2D)}(\epsilon)$. 
With the help of the $\Lop$ operator for \textit{arbitrary layered 
superconductor}, cf. Refs.~\citen{klemm1} and \citen{man1}, these three quantities can be 
generally written as 
\begin{align} 
  \sigma_{ab}(\epsilon) & = {1\over R_{\rm QHE}} 
  \left(\frac{2\tau_0\kb\tc}{\hbar}\right) {N\over s}\Lop {1\over \epsilon}, 
    \label{Lsigma} \\ 
  -\chi_{ab}(\epsilon) & = {\pi\over 3}\mu_0{\kb T\over\Phi_0^2} 
  \xi^2_{ab}(0){N\over s}\Lop {1\over \epsilon}, \label{Lchi} \\ 
  C(\epsilon) & =  {\kb\over 4\pi\xi^2_{ab}(0)}{N \over s} 
                    \Lop {1\over \epsilon}, 
\label{heat} 
\end{align} 
where the ``$ab$''-subscript in \Eref{Lsigma} indicates that the conductivity is 
in the $ab$-planes, while in \Eref{Lchi} it indicates that the vanishing magnetic 
field is  perpendicular to the same planes. 
It is immediately apparent that the common for all these expressions function 
$\Lop \epsilon^{-1}$ cancels when calculating the  $\chi/C,$ $\sigma/C,$ and $\sigma/\chi$ quotients.\cite{Vidal} 
In particular, the temperature independent ratio 
\begin{equation} 
\tau_0={\mu_0\over 3}\xi^2_{ab}(0) 
  \frac{ 
    \sigma_{ab}(\epsilon) 
  }{ 
    -\chi_{ab}(\epsilon) 
  } = \mbox{const} 
\label{tauratio} 
\end{equation} 
provides the best method for probing the time constant $\tau_0$ 
parameterizing the life time of the fluctuation Cooper pairs with zero 
momentum, 
\begin{equation} 
  \tau(\epsilon) = {\tau_0 \over \epsilon}, 
  \qquad 
  \left| \psi_{{\bf p}=0}(t) \right|^2 \propto 
  \exp\!\left( 
    -{t\over\tau(\epsilon)} 
  \right). 
\end{equation} 
The $\tau_0$ constant participates in time-dependent GL (TDGL) theory;
see for example the reviews by Cyrot,\cite{Cyrot} Skocpol and
Tinkham,\cite{skoc} and the textbooks by Abrikosov\cite{Abrikosov} and
Tinkham.\cite{Tinkham} Within the weak coupling BCS theory in the case
of negligible depairing
mechanisms\cite{Abrikosov,Tinkham,Cyrot,abwoo,abtsu,schmid2,maki4,fulde,Yip,abwoo2} $\tau_0$ satisfies the relation
\begin{equation} 
  \frac{\tau^{\rm (BCS)}_{0,\Psi}\kb\tc}{\hbar}={\pi\over 8}, 
    \quad 
  \tau_0^{\rm (BCS)} = \frac{\pi}{16} \frac{\hbar}{\kb\tc}, 
    \quad 
  \tau^{\rm (BCS)}(\epsilon)=\frac{\tau_0^{\rm (BCS)}}{\epsilon} 
  =\frac{\pi}{16} \frac{\hbar}{\kb\tc}{1\over \epsilon}, 
\label{pi8} 
\end{equation} 
where $\tau^{\rm (BCS)}_{0,\Psi} \equiv 2 \tau_0^{\rm (BCS)}$ is the
relaxation time constant for the order parameter being two times
larger.\cite{Tinkham} At the present experimental accuracy this BCS
value agrees well with the experimental data for the layered cuprates.
Thus, the above observation led us to propose the dimensionless
ratio\cite{EPL}
\begin{equation} 
  \tilde \tau_{\rm rel} \equiv \frac{\tau_0}{\tau_0^{\rm (BCS)} } 
  = 32 \frac{\kb\tc\tau_0}{(2\pi\hbar)}= 
  \frac{8\kb\tc\tau_{0,\Psi}} {\pi \hbar }= 
  \frac{16\mu_0}{3\pi\hbar}\xi_{ab}^2(0) 
  \frac{\kb T\sigma_{ab}(\epsilon)}{-\chi_{ab}(\epsilon)} = \mbox{const}, 
\label{taurel} 
\end{equation} 
to be used for more reliable experimental data processing; any deviation 
of $\tilde \tau_{\rm rel}$ from unity should be interpreted as a hint towards 
unconventional behavior and presence of depairing mechanisms. 
Notice also that the BCS value $\pi/8= 0.393$ in 
Eqs.~(\ref{ALformula}), (\ref{pi8}), and (\ref{taurel}) is extremely 
robust, being originally derived for dirty 3D superconductors, and  the 
$\tau_0 \tc$ product remains the same\cite{Yip,varlall} for clean 
2D superconductors and is not affected by the multilaminarity. 
The general formula for the fluctuation conductivity of a 
layered superconductor in perpendicular magnetic field can be also 
rewritten via the layering operator and relative life-time employing 
the 2D results by Redi,\cite{redi} and Abrahams, Prange and 
Stefen\cite{abpran} (APS), cf. also \Rref{manolo}, 
\begin{equation} 
  \sigma_{ab}(\epsilon,h) = \tilde \tau_{\rm rel} 
  \frac{e^2}{16\hbar} {N\over s} \Lop f_{\rm APS} (\epsilon,h), 
\label{APSsigma} 
\end{equation} 
where, for $\epsilon+h > 0$, 
\begin{equation} 
  f_{\rm APS}(\epsilon,h)\equiv {1\over \epsilon} 2 
  \left( 
    \frac{\epsilon}{h} 
  \right)^2 
  \left[ 
    \psi\!\left({1\over 2} + \frac{\epsilon}{2h}\right) 
    - \psi\!\left(1 + \frac{\epsilon}{2h}\right) + \frac{h}{\epsilon} 
  \right], 
\label{fAPS} 
\end{equation} 
is an universal dimensionless function of dimensionless reduced temperature 
$\epsilon$ and dimensionless magnetic field $h.$ The functions 
\begin{equation} 
  \Gamma(z) \equiv \int_0^{\infty} e^{-t}t^{z-1} dt, 
    \quad 
  \psi(z) \equiv \frac{d}{dz}\ln\Gamma(z), 
    \quad 
  \psi^{(1)}(z) \equiv \frac{d}{dz}\psi(z) = \zeta(2,z) 
\end{equation} 
are  respectively the Euler gamma, digamma, and trigamma 
functions. This general formula is often utilized to process the experimental 
data  for the paraconductivity. We provide also several useful 
asymptotics of $f_{\rm APS}(\epsilon,h)$ in different physical conditions, 
\begin{equation} 
  f_{\rm APS}(\epsilon,h)\approx 
  \begin{cases}
    \displaystyle
    {2\over h} \left[1 - \frac{\epsilon}{2h} \ln 2\right], 
                              & h \gg |\epsilon| \\[0.4cm] 
    \displaystyle
    \frac{4}{\epsilon+h}  = 4\;\frac{\tc}{T-T_{c2}(H)}, 
                              & \epsilon +h \ll h \\[0.4cm]  
    \displaystyle
    \left[1 - {1\over2}\left(\frac{h}{\epsilon}\right)^2\right] 
    {1 \over \epsilon} = \left[1-{h^2\over4}\frac{\partial^2}{\partial 
    \epsilon^2}\right] {1\over \epsilon},                   
                              & h \ll \epsilon
     \end{cases}
\label{APSlowh}  
\end{equation} 
For the LD model, for example, the ($h \ll \epsilon$)-asymptotics
gives,\cite{manolo,Dorin,varlall} according to \Eref{LD3},
\begin{equation} 
  \sigma_{ab}(\epsilon,h)\approx \tilde \tau_{\rm rel} \frac{e^2}{16\hbar} 
  {N\over s} 
  \left[ 
    \frac{1}{\sqrt{\epsilon(\epsilon+r)}} -{h^2\over2} 
    \frac{\epsilon(\epsilon+r)+ {3\over 8} r^2}{[\epsilon(\epsilon+r)]^{5/2}} 
  \right]. 
\label{LDmagnetocon} 
\end{equation} 
Note that the classical Aslamazov-Larkin result, \Eref{ALformula}, 
is recovered for $r=0,$ $h=0,$ and $\tilde \tau_{\rm rel}=1.$ 
In the practical application to layered cuprates, however,  we need to take 
into account the nonlocality effects. In $\ecut$-approximation to the 
GL theory we have to subtract the part of the corresponding cutoff area in 
the 2D momentum space. 
Thereby, the fluctuation conductivity is given by the difference 
\begin{equation} 
  \sigma_{ab}(\epsilon,h;c)= \Cop 
  \sigma_{ab}(\epsilon,h)\equiv 
  \sigma_{ab}(\epsilon,h)-\sigma_{ab}(c+\epsilon,h) \approx 
  \sigma_{ab}(\epsilon,h)-\sigma_{ab}(c,h), 
\label{Csigma} 
\end{equation} 
where a cutoff operator $\Cop$ is introduced, and the approximation is 
valid for $\epsilon \ll c.$ Similarly, for the magnetization we have the 
same ``cutoff'' expression which appears when calculating the truncated 
sums over the Landau levels, 
$ 
  \sum_{0}^{n_c-1}=\sum_{0}^{\infty}-\sum_{n_c}^{\infty}, 
$ 
or integral with respect to the dimensionless in-plane kinetic energy, 
\begin{equation} 
  \int_{0}^{c} d{\tilde x}= \int_{0}^{\infty} d{\tilde x} 
  - \int_{c}^{\infty} d{\tilde x}. 
\end{equation} 
 
As a rule the GL theory allows for ultraviolet (UV) regularization---every 
expression can be easily regularized in the local 
($c\rightarrow \infty$)-approximation. 
Therefore the energy cutoff parameter $c$ is not viewed as a tool for UV 
regularization, it is simply an important and immanent parameter 
of the GL theory, being of the order $c\simeq 1.$ 
The cutoff procedure has been essentially introduced 
\textit{from the beginning} in the GL theory.\cite{LLV} 
Unfortunately, for many superconductors systematic 
studies for determination of the energy cutoff parameter are still missing. 
Here we suggest only the simplest possible interpolation formula within 
the LD model for $\epsilon \ll c$, 
\begin{equation} 
  \sigma_{ab}(\epsilon,h) = 
  {\pi\over8} \frac{\tilde\tau_{\rm rel}}{R_{\rm QHE}} 
  {N\over s} \; \Cop f_{\rm LD}(\epsilon,r) 
  \approx \tilde\tau_{\rm rel}\frac{e^2}{16\hbar}{N\over s} 
  \left[ 
    \frac{1}{\sqrt{\epsilon(\epsilon+r)}} -\frac{1}{\sqrt{c(c+r)}} 
  \right], 
\label{cutLD} 
\end{equation} 
which takes into account only the fist nonlocal correction. 
This simple expression fits very well\cite{Rebolledo} the 
experimental data for YBa$_2$Cu$_3$O$_{7-\delta}$ and 
Bi$_2$Sr$_2$CaCu$_2$O$_8.$ 
 
As a last example of the action of the layering operator $\Lop$ we consider 
 the 2D frequency-dependent paraconductivity at zero magnetic field. 
Taking the general expression for $D$-dimensional GL model\cite{Dorsey91} 
and performing carefully the limit $D \rightarrow 2$ 
(note, that there is an omitted term in the expression for the 2D 
conductivity in \Rref{Dorsey91}) we get for the in-plane complex 
conductivity 
\begin{equation} 
  \sigma_{ab}^{*}(\omega\tau(\epsilon))= 
  \sigma'_{ab}\!\left({\omega \tau_0\over \epsilon}\right) +i 
  \sigma''_{ab}\!\left({\omega \tau_0\over 
  \epsilon}\right) = \tilde\tau_{\rm rel}\frac{e^2}{16\hbar}\frac{N}{s} 
  \Lop \left[\frac{1}{\epsilon}\; \varsigma_1\!\left({\omega \tau_0\over 
  \epsilon}\right) + \frac{i}{\epsilon}\; \varsigma_2\!\left({\omega 
  \tau_0\over \epsilon}\right) \right], 
\label{sigmaomega} 
\end{equation} 
or in expanded notation for singe layered superconductor, 
\begin{equation} 
  \sigma_{ab}^{*}(\omega)=\frac{2\tau_0\kb\tc/\hbar}
   {s_{\rm eff} R_{\rm QHE}} 
  \int_{0}^{\pi/2} \frac{d\phi}{\pi/2}\; 
  \frac{\varsigma_1\!\left(\frac{\omega\tau_0}{\epsilon + r\sin^2\phi}\right)
  + i \varsigma_2\!\left({\omega 
      \tau_0\over \epsilon+ r\sin^2\phi}\right)}{\epsilon + r\sin^2\phi}, 
\label{sigmaomegaextended} 
\end{equation} 
where we have for the dimensionless real and imaginary conductivity, 
$\varsigma_1(0)=1,$ 
\begin{align}
  \varsigma_1(z) & \equiv  \frac{2}{z^2} \left[z \arctan(z) - \frac{1}{2} 
                           \ln\!\left( 1 + z^2 \right)\right] 
                  =  \frac{2}{\pi} {\cal P} \int_{0}^{\infty} 
                         \frac{ y\;\varsigma_2(y)}{y^2 - z^2} dy, \\ 
  \varsigma_2(z) & \equiv  \frac{2}{z^2} \left[ \arctan(z) - z + z {1\over 2}
                           \ln\!\left( 1+ z^2 \right) \right] 
                  = - {2 z\over\pi}{\cal P}\int_{0}^{\infty} 
                         \frac{\varsigma_1(y)}{y^2 - z^2} dy. 
\end{align} 
As usual in the above Kramers-Kronig integrations $\cal P$ indicates that the 
principal value of the integral is taken. For computer implementation of 
the $\Lop$ operator we have to verify that, for $r \gg 1,$ 
$\Lop$ is simply equivalent to an incremental operator for 
the spatial dimensionality, 
\begin{equation} 
  \sigma^{\text{(D+1)}}(\epsilon) \approx \Lop\sigma^{\rm (D)}(\epsilon). 
\end{equation} 
 
In many cases the GL results for integer dimensionality are 
well-known and we can derive a generalization for a layered 
system. For both the MT and LD models the integration in \Eref{sigmaomega} can be 
easily programmed, so we have a useful formula for fitting of the ultra high frequency 
measurements of $\sigma_{ab}^{*}(\omega).$ The original explicit expressions 
derived from retarded electromagnetic operator by Aslamazov and 
Varlamov\cite{aslavarla} are too cumbersome to be used by 
experimentalists. Hence, one may realize that the GL theory is not some 
phenomenological alternative to the microscopic BCS theory (this 
scorn, dating back to the beginning of fifties, is still 
living even nowadays among students). The GL theory is a tool 
for applying the theory of superconductivity for the 
important for applications, let us say ``hydrodynamic'', case of low 
frequencies and small wave-vectors. For $\epsilon \ll r$ 
the frequency dependent conductivity 
$\sigma_{ab}^{*}\left(\omega\right)$, having dimension $(\Omega \; 
{\rm cm} )^{-1}$, from \Eref{sigmaomegaextended} displays  3D 
behavior, while in the opposite case of $\epsilon \gg r$ 
it shows 2D character. For thin films of layered superconductors with 
thickness $d_{\rm film}$ we have to calculate the 2D conductivity 
$\sigma^{\rm (2D)} = d_{\rm film}\sigma,$ while for single layered 
films of conventional superconductors, for example, we have to 
substitute in \Eref{sigmaomegaextended} formally  $s_{\rm eff}=d_{\rm film},$ 
and certainly $r=0.$ 
 
 Having analyzed in detail the action of the $\Lop$ operator, 
we developed practically all technical tools necessary to proceed our 
investigation of the thermodynamics of Gaussian fluctuations and 
fluctuation magnetization.

\subsection{Power series for the magnetic moment within the LD model} 
 
\noindent 
We  will calculate in this subsection the nonlinear susceptibility 
by substituting first into the free energy, \Eref{Fmagn}, the heat 
capacity, expressed via the susceptibility from \Eref{chiC}. 
Then, the formula for the magnetization, \Eref{M}, gives 
\begin{equation} 
  \chi(\epsilon,h)= 6\sum_{n=1}^{\infty} (-1)^{n-1} 
  \frac{2n}{\pi^{2n}} \left( 1- \frac{1}{2^{2n-1}} \right) \zeta(2n) 
  h^{2(n-1)} \frac{ \partial^{2(n-1)} } { 
  \partial\epsilon^{2(n-1)} } 
  \chi(\epsilon). 
\end{equation} 
Taking the LD expression for the susceptibility at zero field 
\Eref{chiLD}, calculating the derivatives with respect to 
$\epsilon$ by means of the relation 
\begin{equation} 
  \frac{ \partial^{m} } { 
  \partial\epsilon^{m} } \frac{ 1 } { \sqrt{\epsilon} }= \frac{ 
  (2m-1)!! } { 2^{m} \epsilon^{m} } \frac{ 1 } { \sqrt{\epsilon} }, 
\end{equation} 
and defining the relative susceptibility as 
\begin{equation} 
  {\tilde \chi}_{\rm rel}(\epsilon,h)\equiv 
  \frac{\chi(\epsilon,h)}{\chi(\epsilon)}
  \label{chi} 
\end{equation} 
we obtain 
\begin{align} 
  {\tilde \chi}_{\rm rel}(\epsilon,h;r) =&\,  12 
     \sum_{n=0}^{\infty} \frac{(-1)^n}{2n+1} \left(1- 
     \frac{1}{2^{2n+1}} \right) \frac{(2n+2)!}{2^{2n+1}} \frac{\zeta 
     (2n+2)}{\pi^{2n+2}} \nn \\ 
  &  \qquad \times \left( \frac{h^2}{\epsilon^2} \right)^n 
       \sum_{m=0}^{2n} \frac{(2m-1)!!(4n-2m-1)!!}{m!(2n-m)! 
       (1 +r/\epsilon)^{2n-m}} \nn \\ 
  = &\, 1 - \frac{7}{15} \frac{\epsilon^2 + r\epsilon + 3r^2/8}{(\epsilon+r)^2} 
      \left( \frac{h}{\epsilon} \right)^2 + \cdots . 
\label{chirel} 
\end{align} 
Although these series is found to be a solution to the problem of 
calculating the fluctuational magnetization, 
\begin{equation} 
  M(\epsilon,h) = -\frac{\kb \tc}{\Phi_0 s_{\rm eff}} \, 
  \left\{ 
    \frac{\tilde \chi_{\text{rel}}(\epsilon,h;r)} 
     {6 \sqrt{\epsilon(r + \epsilon)} } 
     - \frac{\tilde \chi_{\text{rel}} (c+\epsilon,h;r)} 
     {6\sqrt{(c+\epsilon)(c + r + \epsilon)}}
 \right\}, 
\label{MlowB} 
\end{equation} 
for the physical conditions of interest, \ie an observable 
effect of magnetic field on the susceptibility, one needs to 
extend the series summation onto arguments $h^2/{\epsilon}^2$ beyond 
the radius of convergence. Analogous series has been already reported 
for the 3D paraconductivity.\cite{Tsuzuki} One of the best devices for 
extending the convergence of series and also for calculating 
slowly convergent series is the 
${\varepsilon}$-algorithm\cite{wynn,shanks} based on Pad\'e 
approximants.\cite{baker} In the next section we describe a 
simplified version of this algorithm suitable for computer implementation.

\subsection{The epsilon algorithm} 
 
\noindent 
The epsilon algorithm is a method for finding the limit $L$ of infinite series 
\begin{equation} 
  L = \lim_{N \rightarrow \infty}S_N 
   \equiv \lim_{N \rightarrow \infty}\sum_{i=0}^N a_i , 
\end{equation} 
in case where only the first $N+1$ terms $a_i$, $i=0,\dots N,$ are known. 
The algorithm operates by employing two rows. The first one, called here 
auxiliary $A$-row, is initially set to zero, \ie 
\begin{equation} 
  A_{0}^{[0]}=0, \qquad A_{1}^{[0]}=0, \qquad 
  A_{2}^{[0]}=0, \dots \qquad A_{N}^{[0]}=0. 
\label{Arowinit} 
\end{equation} 
The second one is sequential $S$-row loaded in zero-order 
approximation with the partial sums of the series 
\begin{equation} 
  S_{0}^{[0]}=a_0, \qquad 
  S_{1}^{[0]}=a_0+a_1,\;\; \dots \qquad 
  S_{N}^{[0]}=a_0+a_1+a_2 + \dots + a_N. 
\label{Srowinit} 
\end{equation} 
The above assignments, as indicated by 
Eqs.~(\ref{Arowinit}) and (\ref{Srowinit}), 
constitute the initialization phase of the $\eps$-algorithm. 
The essence of the latter consists of filling in the so called $\eps$-table 
\begin{align} 
 \begin{pmatrix}
   A_0^{[0]} & A_1^{[0]} & A_2^{[0]} & A_3^{[0]} & \dots\\ 
   S_0^{[0]} & S_1^{[0]} & S_2^{[0]} & S_3^{[0]} & \dots\\ 
   A_0^{[1]} & A_1^{[1]} & A_2^{[1]} & A_3^{[1]} & \dots\\ 
   S_0^{[1]} & S_1^{[1]} & S_2^{[1]} & S_3^{[1]} & \dots\\ 
   \hdotsfor{5}
  \end{pmatrix}
  & = 
  \begin{pmatrix}
   \eps_0^{[0]} & \eps_1^{[0]} & \eps_2^{[0]} & \dots \\ 
   \eps_0^{[1]} & \eps_1^{[1]} & \eps_2^{[1]} & \dots \\ 
   \eps_0^{[2]} & \eps_1^{[2]} & \eps_2^{[2]} & \dots \\ 
   \eps_0^{[3]} & \eps_1^{[3]} & \eps_2^{[3]} & \dots \\ 
   \hdotsfor{4}
  \end{pmatrix} \nn \\
  & = 
  \begin{pmatrix} 
   \hdotsfor{4}                  \\
   \left[ 0/0 \right] & [1/0] & [2/0] & \dots \\ 
   \hdotsfor{4}                  \\ 
   \left[ 1/1 \right] & [2/1] & [3/1] & \dots \\ 
   \hdotsfor{4}                  \\ 
  \end{pmatrix}, 
 \label{etable}
\end{align}
where according to the standard notations $[j/k] = P_j(z)/P_k(z)|_{z=1}$ 
designates a Pad\'e approximant having power $j$ in the nominator 
and, respectively, $k$ in the denominator.\cite{baker}
 
Starting from the $A^{[0]}$- and $S^{[0]}$ rows every subsequent row is 
derived by applying the \textit{cross rule} (known also as the missing 
identity of Frobenius). 
To be specific, for calculation of the $k$th $A$-row we have to solve the 
cross rule equation 
$$ 
  \begin{pmatrix} 
    \dots       & A_{i+1}^{[k-1]}\\ 
    S_i^{[k-1]} & S_{i+1}^{[k-1]}\\ 
    A_{i}^{[k]} & \dots          \\ 
    \hdotsfor{2}
  \end{pmatrix} 
= 
  \begin{pmatrix} 
    \dots       & {\rm North}   \\ 
    {\rm West}  & {\rm East}    \\ 
    {\rm South} & \dots         \\ 
    \hdotsfor{2}
  \end{pmatrix}, 
$$ 
\begin{equation} 
  \left({\rm South}-{\rm North}\right) 
  \left({\rm East}-{\rm West}\right)=1. 
\end{equation} 
Likewise, for calculating the $k$-th $S$ row we have to apply 
the same cross rule 
$$ 
  \begin{pmatrix} 
    \hdotsfor{2} \\ 
    \dots       & S_{i+1}^{[k-1]} \\ 
    A_{i}^{[k]} & A_{i+1}^{[k]}   \\ 
    S_{i}^{[k]} & \dots 
  \end{pmatrix} 
= 
  \begin{pmatrix} 
    \hdotsfor{2}              \\ 
    \dots       & {\rm North} \\ 
    {\rm West}  & {\rm East}  \\ 
    {\rm South} & \dots 
  \end{pmatrix}, 
$$ 
\begin{equation} 
  {\rm South} = 
  {\rm North} + \left( {\rm East} - {\rm West} \right)^{-1}. 
\label{Srow} 
\end{equation} 
Having applied the algorithm we get in the $S$-rows of the 
$\eps$-table, \Eref{etable}, a set of different Pad\'e 
approximants to the limit $L.$ 
The $i$th term of the $k$th $A$-row can be easily obtained by 
\begin{equation} 
  A_i^{[k]}=A_{i+1}^{[k-1]} + 
  \left( S_{i+1}^{[k-1]} - S_{i}^{[k-1]} \right)^{-1}, \qquad 
  \mbox{ for } i=0,1,\dots, N-2k+1, 
\label{Aik} 
\end{equation} 
but for practical implementation of the algorithm, we can omit the index 
of the approximation and to use only one auxiliary row, updating it 
each time, 
\begin{equation} 
  A_i := A_{i+1} + \left( S_{i+1}-S_{i} \right)^{-1}, \qquad 
  \mbox{ for } i=0,1, \dots, N - 2k + 1. 
  \label{Aupdate} 
\end{equation} 
For the $k$-th $S$-row, the $i$-th term reads as 
\begin{equation} 
  S_i^{[k]}=S_{i+1}^{[k-1]}+\left( A_{i+1}^{[k]}-A_{i}^{[k]} \right)^{-1}, 
  \qquad \mbox{ for } i=0,1,\dots, N-2k, 
\label{Srow1} 
\end{equation} 
and can be updated in the same manner as described for the $A$-row, 
\begin{equation} 
  S_i := S_{i+1}+\left( A_{i+1}-A_{i} \right)^{-1}\qquad 
  \mbox{ for } i=0,1,\dots, N-2k. 
\label{Supdate} 
\end{equation} 
 
In order to find an estimate  for  the limit $L$ of the infinite series, 
two different empirical criteria can be implemented. 
In the first one, the $\eps$-table is scanned  for a minimal difference 
$| S_{i+1}^{[k-1]} - S_{i}^{[k-1]} |.$ The limit $L$ is then given by 
\begin{equation} 
 \min_{i,k} \left| S_{i+1}^{[k-1]}-S_{i}^{[k-1]} \right| 
  \Longrightarrow  L \approx S_{i}^{[k-1]}. 
\end{equation}
This minimal difference gives also an estimate for the empirical error 
of the method. 
In the second criterion the $\eps$-table is scanned for the maximum 
of the East $A$-row element, cf Eqs.~(\ref{Srow}) and (\ref{Srow1}), 
\begin{equation} 
  \max_{i,k} \left| A_{i+1}^{[k]} \right| 
  \Longrightarrow L \approx S_{i}^{[k]}. 
\end{equation} 
The reciprocal of the maximum auxiliary value gives in this case 
the estimate for the empirical error of the method. 
It is the second criterion that we have used in the \textsc{fortran90} 
implementation of the $\eps$-algorithm given in Appendix~A. 
Therein we have also made use of pseudo-inverse numbers in order to ensure 
provisions against division by zero in Eqs.~(\ref{Aupdate}) 
and (\ref{Supdate}), 
\begin{equation} 
  z^{-1} : = 
  \begin{cases} 
     0,    &  \text{ for } z  =   0 \\ 
     1/z,  &  \text{ for } z \neq 0 
  \end{cases} . 
\end{equation} 
 
For an illustration, consider the first approximation. In the 
beginning we have for the first $A$-row according to \Eref{Aik} 
\begin{equation} 
  A_0^{[1]} = 
  \left[ 
    \left( a_0 + a_1 \right) - (a_0) 
  \right]^{-1}  = \frac{1}{a_1}, \qquad 
  A_{1}^{[1]}   = \frac{1}{a_2},    \; \dots, \qquad 
  A_{N-1}^{[1]} = \frac{1}{a_N}. 
\end{equation} 
The first $S$-row then reads 
\begin{equation} 
  S_0^{[1]} = a_0 + a_1 + \frac{1}{1/a_2 - 1/a_1},\qquad 
  S_1^{[1]} = a_0 + a_1 + a_2 + \frac{1}{1/a_3 - 1/a_2}, 
  \dots\; , 
\end{equation} 
and for the last element of the $S^{[1]}$-row we have 
\begin{equation} 
  S_{N-2}^{[1]}=a_0 + a_1 +a_2 + \cdots + a_{N-2}+a_{N-1}+(1/a_N - 1/a_{N-1}). 
\end{equation} 
The above approximation $S_{N-2}^{[1]}$ to the limit $L$ is nothing but 
the well-known Aitken's $\Delta^2$-method, which gives an 
\textit{exact result} for the geometric progression 
\begin{equation} 
  S_{N}^{[0]} = 1 + q + q^2 + \cdots + q^{N}, \qquad 
  S_{0}^{[1]} = S_{1}^{[1]}   = S_{2}^{[1]} = \cdots
              = S_{N-2}^{[1]} = \frac{1}{1-q}, 
\end{equation} 
for an arbitrary $q \neq 1.$ This fact can rationalize the success of 
the $\eps$-algorithm when applied to weak magnetic field  series 
expansion of susceptibility. In the 
Euler-MacLaurin summation, Eqs.~(\ref{sum-int}) and (\ref{progression}), 
we have a hidden geometric progression of translation operators. 
 
As a rule divergent series do not exist in physics; 99\% of the
divergent series born by real physical problems can be summed up by
some combination of the Euler-MacLaurin method and the
$\eps$-algorithm and the reason is lies in the analytical dependence
of the coefficients on the index. In the Gaussian spectroscopy of
superconductors, for example, it is necessary series related to
asymptotic expansion of Euler polygamma and Hurwitz zeta functions to
be summed up, but the same methods could be applied to many other
physical problems. The solution often can be derived by less efforts
than required to verify that a series is divergent accordingly some
strict mathematical criterion. Nowadays the mathematical education in
the physical departments is conquered by scholastic mathematicians.
Alas, none of the students of physics knows what really happens when
we press the \fbox{\textsf{sin}} key of a calculator.  On the other
hand this is a commercial secret of the manufacturer.  The physicists
do not even lightly touch the brilliant achievements of mathematics
indispensable not only for the theoretical physics but for
experimentalist to fit their data as well.  This is the motivation why
we, following the spirit of the century of enlightenment, present in
Appendix~A a simple \textsc{fortran90} program illustrating the
operation of the $\eps$-algorithm. Certainly, \textit{fysics is
phun},\cite{Feynman} being in part art \textit{cosa
mentale},\cite{Leonardo} and every new software cannot be foolproof,
but there are methods which must be taken into account in every
complicated calculation.
\pagebreak[2] 
 
\subsection{Power series for differential susceptibility} 
 
\noindent 
Having calculated the relative dimensionless 
susceptibility by employing the $\eps$-algorithm we can recover the 
usual susceptibility from the dimensionless one, 
\begin{equation} 
  \chi(\epsilon,h)={\tilde \chi}_{\rm rel}(\epsilon,h) \;\chi(\epsilon). 
\end{equation} 
In order to take into account the effects of nonlocality 
the cutoff area in the momentum space should  subtracted out 
from the susceptibility 
\begin{equation} 
\chi_{_{\mbox{\ding{36}}}}(\epsilon,h) = \Cop \chi(\epsilon,h) 
 = {\tilde \chi}_{\rm rel}(\epsilon,h) \; 
   \chi(\epsilon) - {\tilde \chi}_{\rm rel}(c+\epsilon,h) \;\chi(c+\epsilon). 
\end{equation} 
Then we can easily find the magnetization 
\begin{equation} 
  M(H,T)=\chi_{_{\mbox{\ding{36}}}}(\epsilon,h) H. 
\end{equation} 
The calculation of the differential susceptibility 
\begin{equation} 
  \chi^{\rm (dif)}(\epsilon,h) = 
  \left(\frac{\partial M}{\partial H}\right)_T, 
\label{difchi} 
\end{equation} 
where $H=H_{c2}(0)h,$ gives an alternative method to determine the
magnetization. Next we define a dimensionless relative differential
susceptibility
\begin{equation} 
   \tilde \kappa \left(\epsilon,h,r\right) 
   \equiv \chi^{\rm (dif)}(\epsilon,h)/\chi(\epsilon). 
\end{equation} 
For this variable, using \Eref{chirel}, we have the series 
\begin{align} 
   \tilde\kappa 
 & = 12 \sum_{n=0}^{\infty} \left(1- \frac{1}{2^{2n+1}} \right) 
  \frac{(2n+2)!}{2^{2n+1}} \frac{\zeta (2n+2)}{\pi^{2n+2}} \left( - 
       \frac{h^2}{\epsilon^2} \right)^n \nn \\ 
 &\quad\qquad \times  \sum_{m=0}^{2n} \frac{(2m-1)!! \; (4n-2m-1)!!} {m! \; 
       (2n-m)! \;(1+r/\epsilon)^{2n-m}} \nn \\ 
 & =  1-{7\over 5} \frac{\epsilon^2 + r\epsilon +3r^2/8}{(\epsilon+r)^2} 
   \left( \frac{h}{\epsilon} \right)^2 + \dots, 
\label{kappasum} 
\end{align} 
which, just as done in deriving \Eref{chirel}, can be summed up by 
means of the $\eps$-algorithm. For instance, in the local GL limit we 
have for the magnetization
\begin{align} 
  M = \int_0^{H} \chi^{\rm (dif)}(T,H) dH 
    &= \chi(\epsilon)H_{c2}(0) \; \int_0^h 
   \tilde \kappa (\epsilon,h') dh' \nn\\
   & = \chi(\epsilon)H_{c2}(0) \; {\tilde \chi}_{\rm rel} (\epsilon,h) h. 
\end{align} 
For the further analysis, however, it is more suitable to introduce a 
dimensionless magnetization 
\begin{equation} 
  {\tilde m} \equiv - {M \over M_{0}} =-\frac{\Phi_0}{\kb \tc} 
  {s \over N} \;M, \qquad 
  M_0 \equiv \frac{\kb \tc}{\Phi_0} \frac{N}{s}. 
\end{equation} 
Then, using the relation 
\begin{equation} 
  {\tilde \chi}_{\rm rel}(\epsilon,h) = 
  {1\over h} \int_0^h {\tilde \kappa} (\epsilon,h') dh' 
\end{equation} 
the result for the dimensionless fluctuation magnetization takes the form 
\begin{equation} 
  {\tilde m} (\epsilon,h)= 
  {1\over 6} \, {1 \over \sqrt{\epsilon(\epsilon+r)}}\, 
  {\tilde \chi}_{\rm rel}(\epsilon,h) h 
 = {1 \over 6} \, {1 \over \sqrt{\epsilon(\epsilon + r)}}\, 
  \int_0^h  {\tilde \kappa} (\epsilon,h') dh'. 
\label{weakH} 
\end{equation} 
 
In this section we have calculated the magnetization by means of power
series in the magnetic field assuming in the beginning $h/\epsilon\ll
1.$ In the next section we develop another method for calculating the
fluctuation magnetic moment which is appropriate for strong magnetic
fields and allows for studying the high magnetic field asymptotics for
large enough values of the reduced magnetic field, $h/\epsilon\gg 1.$
The overlap between these expansions about $h/\epsilon\simeq 1$ would
be a test for the accuracy of the calculations.

\section{Strong magnetic fields} 
 
\subsection{General formula for the free energy} 
 
\noindent 
In order to derive a general formula for the Gibbs free energy for 
arbitrary non-vanishing magnetic field we will start again by 
representing the free energy  density as a sum over the energy spectrum, 
\Eref{Feh}, 
\begin{equation} 
  F(\epsilon,h) = F_0 \; 2h\, 
  \Lop \sum_{n=0}^{n_c-1}\left[ 
  \ln\!\left(n + {1 \over 2}+\frac{\epsilon}{2h}\right) + \ln(2h) \right], 
\label{GFE} 
\end{equation} 
where 
\begin{equation} 
  F_0 \equiv\frac{\kb\tc}{4\pi\xi_{ab}^2(0)}{N\over s} 
  = {1 \over 2} M_0 B_{c2}(0). 
\label{F0} 
\end{equation} 
The first way to go in deriving convenient for programming formula 
is to calculate the action of the $\Lop$ operator on the 
integrand, cf. \Rref{Baraduc}. In this case we write down the 
free energy as a finite sum over the Landau levels 
\begin{equation} 
  F(\epsilon,h) = \frac{\kb\tc}{4\pi\xi_{ab}^2(0)}{N\over s} 
  \; 2h \sum_{n=0}^{n_c-1} \Lop\ln\!\left[\epsilon+h(2n+1)\right], 
\label{sumln} 
\end{equation} 
where, according to \Eref{LDln}, 
\begin{equation} 
  \Lop^{\rm (LD)} \ln\!\left[\epsilon+h(2n+1)\right] = 
  2\ln\frac{\sqrt{\epsilon+h(2n+1)} + \sqrt{\epsilon + h(2n+1) + r}}{2}. 
\end{equation} 
This formula is useful especially in the case of strong 
magnetic fields when the finite series are not too long. However, 
in order to have a good working expression, applicable to all cases, it 
is much better to solve the problem analytically. Towards this end 
consider the last term in the integrand of \Eref{GFE}. The summation 
of this constant term and simply yields the cutoff parameter $c$ 
\begin{equation} 
  2h\sum_{n=0}^{n_c-1}1=(2h)n_c=c. 
\end{equation} 
Next we introduce a dimensionless function 
\begin{equation} 
  x(\epsilon,h)\equiv{1\over2}+ {\epsilon\over 2h} 
  = \frac{\epsilon+h}{2h} 
  = {1\over2H}\left(T-T_{c2}(H)\right) 
  \left. 
  \left(-\frac{\partial H_{c2}(T)}{\partial T}\right)\right|_{T=\tc-0}, 
\label{xeh} 
\end{equation} 
which is the argument of some of the analytical functions we 
use in the following. Further, we have to present the sum in 
\Eref{GFE} as a difference of two appropriately regularized 
infinite series 
\begin{equation} 
 \sum_{n=0}^{n_c-1} \ln(n+x)= 
  \Reg{\zeta}\sum_{n=0}^{\infty}\ln(n+x) 
  -\Reg{\zeta}\sum_{n_c}^{\infty} \ln(n+x). 
\end{equation} 
In fact, one does not have any other possibility except the
$\zeta$-regularization
\begin{equation} 
  -\Reg{\zeta}\sum_{n=0}^{\infty} \ln(n+z) 
  = \left.\frac{\partial}{\partial \nu} \zeta(\nu,z)\right|_{\nu=0} 
  =\ln \frac{\Gamma(z)}{\sqrt{2\pi}}, 
\label{sumofln} 
\end{equation} 
based on one relation between the Euler $\Gamma$-function and the
Hurwitz $\zeta$-function, and the definition of the logarithmic
function
\begin{equation} 
  \ln z = \lim_{\nu\rightarrow 0}\frac{z^{\nu}-1}{\nu} 
        = \left. \frac{\partial}{\partial \nu} z^{\nu}\right|_{\nu=0}. 
\end{equation} 
According to the famous results by Riemann,  the analytical 
continuations of the $\zeta$-function and the factorial $n!$ are 
unique. Therefore the UV regularization of the partition function in 
the GL model in an external magnetic field is practically included in the 
second, Gauss definition of the $\Gamma$-function as a infinite product, 
see \eg \Rref{bronsh}, 
\begin{equation} 
  \int_{0}^{\infty} t^{z-1} {\rm e}^{-t} dt 
  \equiv \Gamma(z) \equiv 
  \lim_{n_c \rightarrow \infty}
  \frac{n_c! \; n_c^{z-1}}{z(z+1)(z+2)\dots (z+n_c-1)}. 
\label{Euler} 
\end{equation} 
Let us recall some particular values, 
\begin{equation} 
  \Gamma(n+1)=n!, \qquad 
  \Gamma(1)=0!=1, \qquad 
  \Gamma\!\left({1\over2}\right)= \sqrt\pi, 
\label{Gamma12} 
\end{equation} 
and the Stirling's approximation for $n_c \gg 1,$ derived by 
Gaussian saddle point approximation applied to the first, Euler
definition of the $\Gamma$-function, \Eref{Euler}, 
\begin{equation} 
  n_c! \approx \left(\frac{n_c}{\rm e}\right)^{n_c} \sqrt{2\pi n_c}, 
  \qquad 
  \ln \left(n_c!\right) 
  \approx\left(n_c+{1\over2}\right)\ln n_c-n_c+\ln\sqrt{2\pi}. 
\end{equation} 
For the local limit or for the case of weak magnetic 
fields we shall also make use of the asymptotic formulae for $z\gg 1$ 
\begin{align} 
  \ln \Gamma(z)  &\approx \left(z-{1\over2}\right)\ln z - z 
                          +{1\over2}\ln(2\pi)+{1\over 12z}, \\
  \psi^{(-1)}(z) &\approx\left(z-{1\over2}\right)\ln z-z+{1\over 12z}, 
                                               \label{psim} \\
  \psi(z)        &\approx\ln z - \frac{1}{2z} - \frac{1}{12 z^2}, \qquad 
                  \psi^{(1)}(z) \equiv \zeta(2,z) \approx \frac{1}{z}
                  + \frac{1}{2z^2} + \frac{1}{6z^3}. 
  \label{psiasy} 
\end{align} 
Substituting the Stirling asymptotics in the second Gauss definition,
\Eref{Euler}, and taking a logarithm we arrive at the function
$\psi^{(-1)}(z)$, generating the polygamma functions
\begin{align} 
  \psi^{(-1)}(z) &\equiv \lim_{n_c \rightarrow \infty} \left\{ 
                  -\sum_{n=0}^{n_c-1}\ln(n+z)+\left(n_c-{1\over2}+z\right) 
                  \ln\left(n_c\right)-n_c \right\} \nn \\
                 &= \ln \frac{\Gamma(z)}{\sqrt{2\pi}}. 
\end{align}
As a result, the above Gauss definition for $\ln\Gamma(z)$ solves
the problem for UV regularization of the infinite sum of logarithms,
\Eref{sumofln}. The first derivative of this equation gives the
well-known definition of the digamma function $\psi(z)\equiv
\psi^{(0)}(z),$
\begin{equation} 
  \psi^{(0)}(z)\equiv\frac{d}{dz}\psi^{(-1)}(z)= 
     -\Reg{\zeta} \sum_{n=0}^{\infty}{1\over n+z} 
   = \lim_{n_c \rightarrow \infty} 
     \left\{ 
       -\sum_{n=0}^{n_c-1}\frac{1}{n+z} +\ln\left(n_c\right) 
     \right\}. 
\end{equation} 
In particular, 
\begin{equation} 
  -\psi(1)=C_{\rm Euler} =\lim_{n_c \rightarrow \infty} \left\{ 
  \sum_{n=1}^{n_c-1}{1\over n} -\ln\left(n_c\right) \right\} 
  =0.577216\dots \;\; . 
\end{equation} 
All other polygamma functions are actually Hurwitz $\zeta$-functions 
with integer first argument $\ge 2$ and the sums are trivially convergent, 
\begin{equation} 
  \psi^{(N)}(z)=\frac{d^N}{dz^N}\psi(z) = (-1)^N N!\;\zeta(N+1,z). 
\end{equation} 
To summarize, we have applied the well known
$\zeta$-technique\cite{zeta} for UV regularization of the partition
function and revealed that the archetype of this powerful method comes
from the century of enlightenment and finally we can bring the free
energy, \Eref{GFE}, to the form
\begin{align} 
  F(\epsilon,h) & = {T\over \tc}F_0 \; \left\{2h\, \Lop 
       \left[-\ln\Gamma\!\left({1\over2}+\frac{\epsilon}{2h}\right) 
       +\ln\Gamma\!\left({1\over2}+\frac{\epsilon+c}{2h}\right) 
       \right] + c\ln(2h)\right\} \nn \\ 
  & =  \frac{\kb T}{4 \pi \xi^2_{ab}(0)}{N\over s} \; 
       \Lop \Cop \left[-(2h)\ln \frac{\Gamma\!\left({\epsilon+h\over 2h 
       }\right)}{\sqrt{2\pi}} - \epsilon \ln(2h) \right]. 
\label{theresult} 
\end{align} 
For weak magnetic field, $h\ll \epsilon,$ cf. Eqs.~(\ref{LDxlnx}),
(\ref{LDln}), (\ref{psim}), and (\ref{weakfieldM}),
\begin{equation} 
  - (2h)\ln \frac{\Gamma\!\left({\epsilon+h\over 2h }\right)}{\sqrt{2\pi}} 
  - \epsilon \ln(2h) \approx -\epsilon 
  \left[\ln(\epsilon) - 1 \right] + {1\over 6}{h^2\over\epsilon}. 
\label{weakfieldM1} 
\end{equation} 
This is our main analytical result and all thermodynamic 
properties now can be obtained via derivatives. However, having this 
analytical result it is trivially to check that it can be derived 
by finite sums. The latter do not require UV regularization and the Euler 
$\Gamma$-function is commonly available in many textbooks on mathematical 
analysis.

\subsection{Fluctuation part of thermodynamic variables} 
 
\noindent 
Having an analytical result for the free energy we can 
easily find other thermodynamic variables by differentiating. 
The magnetization, for example, is given by the derivative 
\begin{equation} 
  M =-\left(\frac{\partial F}{\partial B}\right)_T 
    =-\frac{1}{B_{c2}(0)}\left(\frac{\partial F}{\partial h}\right)_{\epsilon} 
    =-M_0 \;{\tilde m}, 
\end{equation} 
where a dimensionless diamagnetic moment is introduced 
\begin{align} 
 {\tilde m}(\epsilon,h)  \equiv  -\frac{M}{M_0}
   &= {c\over 2h} - \Lop \left[ \ln\Gamma\!\left(\frac{\epsilon+h}{2h}\right) 
      -\ln\Gamma\!\left(\frac{\epsilon+c+h}{2h}\right)\right] \nn \\ 
   & \quad + \Lop \left[ 
     \frac{\epsilon}{2h}~\psi\!\left(\frac{\epsilon+h}{2h}\right) 
     -\frac{\epsilon+c}{2h}~\psi\!\left(\frac{\epsilon+h+c}{2h}\right)\right]. 
\label{final} 
\end{align} 
In expanded notations within the  LD model this formula, 
according to Eqs.~(\ref{LDparameter}), (\ref{seff}), (\ref{omega1}), 
 (\ref{M0}), and (\ref{layerLD}),  reads as 
\begin{align} 
  M^{\rm (LD)}(\epsilon,h)  = & -\frac{\kb\tc}{\Phi_0 s_{\rm eff}} \,
   \Biggl( {c\over 2h} + {2\over\pi} \int_0^{{\pi\over2}} d\phi 
   \left\{-\left[ 
     \ln \Gamma\!\left(\frac{\epsilon+r\sin^2\phi+h}{2h}\right) 
   \right. \right. \nn \\ 
 & -  \left.\ln\Gamma\!\left(\frac{c+\epsilon+r\sin^2\phi + h}{2h}\right) 
      \right] + \left[ \frac{\epsilon+r\sin^2\phi}{2h} \,
        \psi\!\left(\frac{\epsilon+r\sin^2\phi + h}{2h}\right)\right.  \nn \\ 
 & -  \left.\left. 
  \frac{c+\epsilon+r\sin^2\phi}{2h} \,
        \psi\!\left(\frac{c+\epsilon+r\sin^2\phi+h}{2h}\right) 
    \right] 
  \right\} 
\Biggl), 
\label{LCM} 
\end{align} 
where $\phi={1\over2}\theta.$ For $|\epsilon|, h \ll r, c$ this
general expression recovers the local 3D result, \Eref{zetaM},
analyzed later in Sec.~3.4, while in the opposite case of extremely
high anisotropy $r<|\epsilon|,h\ll c$ we get the local 2D result,
\Eref{localmagn}. Here we want to emphasize the existence to mention a
universal magnetization law at $T=\tc,$ or $\epsilon=0,$ which can be
observed for many high-$\tc$ materials at strong magnetic fields $h\gg
r$
\begin{equation} 
  -M(\tc,B)\,{\Phi_0 s_{\rm eff} \over  \kb\tc} 
   = \tilde m= {1\over 2}\ln 2\; 
      U_M\!\left({2\over c}{B\over B_{c2}(0)}\right), 
\label{UMagn} 
\end{equation} 
where the universal function of the nonlocal magnetization 
\begin{equation} 
  U_M(y)\equiv \frac{2}{\ln 2} \left\{\ln \Gamma\!\left({1\over y} 
  + {1\over 2}\right)-{1\over 2}\ln\pi + {1\over y}
  \left[1-\psi\!\left({1\over y}+ {1\over 2}\right)\right]  \right\} 
\label{UM} 
\end{equation} 
is normalized so that $U_M(0)=1,$ $U_M(\infty)=0,$ $y\equiv 2h/c.$ For
conventional bulk superconductors the nonlocality effects on
magnetization are well understood, see for example
Refs.~\citen{gollub1,patton1,leepay,kurk,maki2,maki3}.  To the best of
our knowledge, the first observation of fluctuation-induced
diamagnetism for a cuprate superconductor well inside the
finite-magnetic-field regime was reported by Carretta
\etal\cite{Carretta:00} for YBa$_2$Cu$_3$O$_{6+x}.$ Soon after,
analogous measurement was reported for La$_{1.9}$Sr$_{0.1}$CuO$_4$ by
Carballeira \etal\cite{Carballeira:00} Being familiar with the
preliminary version of the present review (cf. \Rref{review})
Carballeira \etal\ have entirely based their interpretation and
theoretical analysis on \Eref{LCM} and \Eref{localmagn} below.  Alas,
we find it very disappointing and impolite that the authors of
\Rref{Carballeira:00} do not give any credits (\eg in the author list,
acknowledgments, or references section) to the author (the first
author of the present review, T.~M.) of the theory they have used. We
will not discuss in any detail their attitude and would instead refer
to the Comment.\cite{comment}
 
Returning now to the general expression for the magnetization,
\Eref{final}, we derive another expression for the relative
differential susceptibility based on \Eref{weakH}
\begin{align} 
  \kappa(\epsilon,h) = &\, 6\sqrt{\epsilon(\epsilon+r)}\; \left( 
  \frac{\partial \tilde m}{\partial h} \right) \nn \\ 
 = &\, 
   6\sqrt{\epsilon(\epsilon+r)}\; 
   \Lop \biggl[ -\frac{c}{2h^2} - 
  \frac{\epsilon^2}{4h^3}~\psi^{(1)}\!\left(\frac{\epsilon+h}{2h}\right)\nn \\ 
  & +\frac{\left(\epsilon+c\right)^2}{4h^3}\,
     \psi^{(1)}\!\left(\frac{\epsilon+h+c}{2h}\right) \biggr]. 
\label{kappacut} 
\end{align} 
The comparison of this result with \Eref{kappasum} is one 
of the best methods to check the accuracy of the programmed formulae. 
Analogously, differentiating the free energy with respect to the 
temperature $T=(1+\epsilon)\tc$ we derive the general formula for 
the most singular part of the entropy (neglecting the derivative 
of the $T$-prefactor in \Eref{theresult}), 
\begin{equation} 
  S \equiv -{1\over \tc}\frac{\partial F}{\partial \epsilon}= 
  \frac{\kb}{4\pi\xi_{ab}^2(0)} \frac{N}{s}\; 
  \Lop \left[ 
  \psi\!\left(\frac{\epsilon+h}{2h}\right)- 
  \psi\!\left(\frac{\epsilon+h+c}{2h}\right) \right], 
\label{entropy1} 
\end{equation} 
and the most singular part of the heat capacity 
\begin{align} 
 C(\epsilon,h) & = \frac{\partial S}{\partial 
   \epsilon} = -{1\over \tc} \frac{\partial^2 F}{\partial \epsilon^2} \nn \\ 
 & = \frac{\kb}{4\pi\xi_{ab}^2(0)} 
     \frac{N}{s}\; \frac{1}{2h} 
     \Lop\left[ 
       \psi^{(1)}\!\left(\frac{\epsilon+h}{2h}\right)
       - \psi^{(1)}\!\left(\frac{\epsilon+h+c}{2h}\right) 
     \right]. 
\label{Ccut} 
\end{align} 
This expression for $C$ can be directly derived from the starting
formulae (\ref{Feh}) and (\ref{GFE}).  The sums for the heat capacity
are convergent, cf. \Rref{LLV}, and do not require any regularization.
The simplest way to reproduce the analytical result for the free
energy density, \Eref{theresult}, is to integrate two times the result
for its second derivative, \ie that for the heat capacity,
cf. \Rref{todor}. In general finite sums from $0$ to $n_c - 1$ for
logarithms and powers can be found in many textbooks on mathematics and
all our results can thus be easily checked even by experimentalists.
 
The fluctuation part of the entropy $S$ is proportional to the mean
square of the order parameter $\Psi$, \ie the volume density of
fluctuation Cooper pairs. The thermally averaged density deserves a
special attention because it is the main ingredient of the
self-consistent treatment of the interaction of order parameter
fluctuations. This Hartree type approximation due to Ullah and
Dorsey\cite{ulah} will be briefly described in the next subsection.
 
In the following, for completeness, we will derive the local 2D
asymptotics applicable for $|\epsilon|, h \ll c.$ The substitution of
the first term from \Eref{psim} into the general formula for the free
energy, \Eref{theresult}, gives
\begin{equation} 
 \tilde f_{\rm 2D}\equiv {F(\epsilon,h)\over F_0}
 = -(2h)\, \psi^{(-1)}\left(\frac{\epsilon+h}{2h}\right)-\epsilon\ln(2h)+ 
  A(c)\epsilon + B(c) + O({1/c}), 
\label{localresult} 
\end{equation} 
where for $\epsilon \ll c$, cf. \Eref{Fe}, 
\begin{equation} 
  f_c(\epsilon) \equiv A(c)\epsilon + B(c) 
  \approx \int_0^{c+\epsilon} \ln \tilde x\; d{\tilde x}
  \approx \epsilon \ln c + c(\ln c -1) 
\label{linfunc}. 
\end{equation} 
This irrelevant for the fluctuation phenomena linear function of $\epsilon$ 
gives constant additions to the free energy $F_c=F_0 B(c),$ and 
entropy $S_c= - F_0 A(c)/\tc$ and can be omitted hereafter. 
The subtraction of $F_0 f_c$ from the free energy, \Eref{theresult}, 
 can be considered as a cutoff procedure for UV regularization, 
\begin{equation} 
  \Reg{\mbox{\ding{36}}} F(\epsilon,h) = F(\epsilon,h) - 
  (F_c -\tc S_c\epsilon), 
\label{LinUV} 
\end{equation} 
which, when applied, allows the analysis of the local GL approximation 
to be carried out simply as $(c 
\rightarrow \infty)$-limit. Now a trivial differentiation gives 
for the dimensionless magnetization, being a positive quantity, 
\begin{align} 
  \tilde m_{\rm 2D}(\epsilon,h) ={-M(\epsilon,h)\over M_0} 
  &= \frac{\epsilon}{2h} \left[ 
     \psi\!\left(\frac{\epsilon}{2h}+{1\over2}\right) -1\right]- 
     \psi^{(-1)}\!\left(\frac{\epsilon+h}{2h}\right) \nn \\
  &= {1\over2}\;\frac{\partial \tilde f_{\rm 2D}}{\partial h} 
\label{localmagn} 
\end{align} 
This result is also a local $(c\rightarrow \infty)$-asymptotic of 
\Eref{final}, which for $h\ll\epsilon$ yields 
\begin{equation} 
  \tilde m_{\rm 2D}\approx h/6\epsilon . 
\label{weakfieldM} 
\end{equation} 
In the general case the local approximation gives $\tilde m= \Lop \tilde m_{\rm 2D},$ or for the LD model 
\begin{align} 
  \tilde m(\epsilon,h;r) = -\frac{M}{M_0} 
   = & \int_{0}^{\pi/2} 
       \frac{d\phi}{\pi/2} \left\{ \frac{\epsilon+r\sin^2 
       \phi}{2h}\left[\psi\!\left(\frac{\epsilon+r\sin^2 
       \phi}{2h}+{1\over2}\right) -1\right]\right. \nn \\ 
    &  \qquad\qquad\qquad \left. -\ln \Gamma\!\left(\frac{\epsilon+r\sin^2 
       \phi}{2h}+{1\over2} \right) + {1\over2}\ln(2\pi)\right\}. 
\end{align} 
The next differentiation with respect to the magnetic field, using 
Eqs.~(\ref{chiLD}), (\ref{difchi}),  gives the relative dimensionless 
susceptibility 
\begin{equation} 
  \tilde \kappa_{\rm 2D}(\epsilon,h) = 
    6\epsilon\left(\frac{\partial \tilde m}{\partial h}\right) 
  = 12\left(\frac{\epsilon}{2h}\right)^2\left[1-\frac{\epsilon}{2h} 
    \psi^{(1)}\!\left(\frac{\epsilon}{2h}+{1\over2}\right)\right] 
  = {\chi^{\rm (dif)}(\epsilon,h)\over \chi(\epsilon)}, 
\label{localkappa} 
\end{equation} 
which is also a local $c \gg h, |\epsilon|$ asymptotic of \Eref{kappacut}. 
For the LD model after averaging with respect to the Josephson phase, 
according to \Eref{omega1}, we obtain 
\begin{equation} 
  \kappa(\epsilon,h;r)= 12 \int_{0}^{\pi/2} \frac{d\phi}{\pi/2} 
  \left(\frac{\epsilon+r\sin^2 \phi}{2h}\right)^2 
  \left[1 -\frac{\epsilon+r\sin^2 \phi}{2h} 
  \zeta\!\left(2,\frac{\epsilon+r\sin^2 \phi}{2h}\right) \right]. 
\end{equation} 
This final result can be directly compared to low field series 
expansion Eq.~(\ref{kappasum}). 
Similar differentiations of the free energy, \Eref{localresult}, 
with respect to the temperatures gives the most singular part of 
the entropy 
\begin{equation} 
  \tilde s_{\rm 2D}\equiv -\frac{\partial}{\partial\epsilon} \tilde f_{\rm 2D} 
  = \left[ \psi\!\left(\frac{\epsilon+h}{2h}\right)+\ln(2h) \right] 
  = \tc S(\epsilon,h)/F_0, 
\end{equation} 
and of the heat capacity 
\begin{equation} 
  \Lop\, \tilde c_{\rm 2D}\equiv - \Lop 
  \frac{\partial^2}{\partial \epsilon^2} \tilde f_{\rm 2D}(\epsilon,h) = 
  {1\over 2h}\psi^{(1)}\left(\frac{\epsilon+h}{2h}\right) 
  = \tc C(\epsilon,h)/F_0. 
\label{localC} 
\end{equation} 
Restoring the $T$ prefactor instead of $\tc$ in \Eref{Fe}, as was done
 in \Eref{theresult}, we arrive at a slightly different expression for
 the fluctuation part of the free energy $F= F_0 (1+\epsilon) \tilde
 f_{\rm 2D} (\epsilon,h)$ and the heat capacity
\begin{align} 
  \Lop \, \tilde c_{\rm 2D} & =  - (1 + \epsilon) \frac{\partial^2}{\partial 
  \epsilon^2} (1+\epsilon)\Lop\,\tilde f_{\rm 2D}(\epsilon,h) \nn \\ 
  & = (1+\epsilon)\left[(1+\epsilon)\Lop\, \tilde c_{\rm 2D} 
     + 2 \Lop \,\tilde s_{\rm 2D} \right] =\frac{\tc C(\epsilon,h)}{F_0}, 
\end{align} 
which gives 
\begin{equation} 
  C(\epsilon,h)=\frac{(1+\epsilon)\kb}{4\pi 
  \xi_{ab}^2(0)}{N\over s} 
  \Lop \left[ {1+\epsilon \over 2h} 
    \psi^{(1)}\!\left({\epsilon + h \over 2h}\right) + 
   2\psi^{(0)}\!\left({\epsilon + h \over 2h}\right) + 2\ln(2h) 
  \right]. 
\label{tedious} 
\end{equation} 
For zero magnetic field we have 
\begin{equation} 
  C(\epsilon,h=0)=\frac{(1+\epsilon)\kb}{4\pi 
  \xi_{ab}^2(0)}{N\over s}\left[ 
  (1 + \epsilon)\, \Lop\, {1\over \epsilon} 
     -2\epsilon\, \Lop \, \ln{1\over\epsilon}\right], \end{equation} 
which in the LD model takes the form 
\begin{equation} 
  C(\epsilon,r)=\frac{(1+\epsilon)\kb}{4\pi 
  \xi_{ab}^2(0)}{N\over s}\left[\frac{(1+\epsilon)} 
  {\sqrt{\epsilon(r+\epsilon)}}-2\epsilon \; 2 
  \ln{2\over\sqrt{\epsilon}+\sqrt{r+\epsilon}}\right]. 
\label{Crplus} 
\end{equation} 
These expression differs from Eqs.~(\ref{CLD}), and (\ref{heat}).
However, the $(1+\epsilon)^2\approx 1+ 2\epsilon$ correction and the
less singular part of the heat capacity $2(1+\epsilon)F_0 \Lop \tilde
s_{\rm 2D}/\tc,$ which appears due to differentiation of $T$ in the
numerator of \Eref{Fe} and \Eref{theresult}, are difficult to be
identified experimentally.
 
For the superconducting phase below the critical temperature,
$0 < -\epsilon\ll 1,$ one has to take into account more or less space
homogeneous order parameter $\Psi_\epsilon$ which minimizes the
nongradient part of the free energy density $F=a(\epsilon)n_\epsilon +
{1\over2}n_\epsilon^2,$
\begin{equation} 
  \Psi_{\epsilon}=\sqrt{a_0(-\epsilon)/b},\qquad 
  n_{\epsilon}=\Psi_{\epsilon}^2=a_0(-\epsilon)/b. 
\end{equation} 
The fluctuations around this minimum 
\begin{equation} 
  \Psi = \Psi_{\epsilon}+\Psi^{\prime}+i\Psi^{\prime\prime}, \qquad 
  n    = \Psi^2= n_{\epsilon}+ 
 2\Psi_{\epsilon}\Psi^{\prime}+ 
 \left(\Psi^{\prime}\right)^2+\left(\Psi^{\prime\prime}\right)^2 
\end{equation} 
should be considered as a small perturbation, thus only  the  quadratic term 
in the free energy is taken into account, 
\begin{equation} 
  F(\epsilon<0)= a(\epsilon) + {1\over2} b n^2 \approx 
  -{1\over2b}a_0^2\epsilon^2 + 
  a_0(-2\epsilon)\left[1\left(\Psi^{\prime}\right)^2 + 
  0\left(\Psi^{\prime\prime}\right)^2 \right]. 
\label{below} 
\end{equation} 
The first term in this equation corresponds to the jump in the 
heat capacity $\Delta C = a_0^2/b\tc$ at $\tc.$ The linear term 
$\propto\Psi^{\prime}$ simply cancels. The phase 
fluctuations $\propto\left(\Psi^{\prime\prime}\right)^2$ are 
coupled to the plasmons and vortexes but they are irrelevant for 
the thermodynamic fluctuations significantly below $\tc.$ In this 
way mainly  fluctuations related to the modulus of the order parameter are 
essential for the heat capacity below $\tc.$ Finally, the 
comparison of the second term $\propto\Psi^{\prime}$ in 
\Eref{below} with the corresponding expression above $\tc$ 
\begin{equation} 
  F(\epsilon>0)= a(\epsilon) + {1\over2} b n^2 \approx 
  a_0(\epsilon)\left[\left(\Psi^{\prime}\right)^2 + 
  \left(\Psi^{\prime\prime}\right)^2 \right], 
\end{equation} 
provides a prescription to derive the fluctuation part below 
$\tc$ from the fluctuation expression for the normal phase above $\tc$ 
\begin{equation} 
  {1\over2}\, \Lop \, {1\over (-2\epsilon)} 
  \leftarrow 
  \Lop\, {1\over \epsilon}. 
\end{equation} 
Applying this prescription to \Eref{Crplus} results in the following
expression
\begin{equation} 
  C(\epsilon<0,r)=\frac{(1+\epsilon)\kb}{4\pi 
  \xi_{ab}^2(0)}{N\over s} \; {1\over2} 
  \left[ 
    \frac{(1+\epsilon)} {\sqrt{(-2\epsilon)(r-2\epsilon)}}-2\epsilon \; 
    2 \ln{2\over\sqrt{(-2\epsilon)}+\sqrt{r-2\epsilon}} 
  \right]. 
\label{Crminus} 
\end{equation} 
These fluctuation part as well as the the phonon heat capacity should
be subtracted from the experimental data in order to extract the jump
$\Delta C$ and related to it penetration depth $\lambda_{ab}(0).$ Such
a procedure, in fact, gives a purely thermodynamic method to determine
the latter quantity.
 
The dimensionless functions Eqs.~(\ref{localresult}), (\ref{localmagn}), 
(\ref{localkappa}), and (\ref{localC})) derived with the local approximation 
are just as important for the thermodynamics of the 
layered superconductors as is the APS function for the paraconductivity, 
\Eref{fAPS}. The operator $\Lop$ gives the 
possibility to extend the 2D analytical result for layered or even 
isotropic 3D superconductor. Additionally the $\Cop$ operator 
gives the energy cutoff approximation for the 
nonlocality effects in the conducting \cuo planes. Therefore the 
analytical 2D result plays a key role for the fluctuation phenomena 
in layered superconductors. 
 
We will finish the analysis of the 
local $c\gg |\epsilon|, h$ 2D approximation $h\gg r,$ \ie 
\begin{equation} 
 \mu_0 H \gg r B_{c2}(0) = 
  \left(\frac{2 \xi_c(0) N}{s}\right)^2 \frac{\Phi_0}{2\pi\xi^2_{ab}(0)}, 
\end{equation} 
with the important case of strong magnetic field $h \gg |\epsilon|.$ 
Under these conditions ($|\epsilon|,r\ll h \ll c$) the layered 
superconductors display a magnetization corresponding to the local 2D one in 
strong magnetic fields. The substitution of $\epsilon=0$ in 
\Eref{localmagn}, using \Eref{Gamma12}, recovers the result 
by Klemm, Beasley, and Luther\cite{klemm2} 
\begin{equation} 
  {\tilde m}(h \gg r,\epsilon \ll h) \approx 
  0.3465735902799726\dots, \qquad 
  -M \approx {\ln 2\over 2} \;\frac{\kb \tc}{\Phi_0} {N\over s}. 
\label{KlemmBeasleyLuther} 
\end{equation} 
In the concise review by Koshelev\cite{kosh} on the properties of 2D GL model 
the calculation of ${1\over 2}\ln 2\approx 0.346$ by infinite series with three
decimal digits accuracy is described in great details. 

\subsection{Self-consistent approximation for the LD model} 
 
\noindent 
The bulk (3D) density of the fluctuation Cooper pairs 
$n(\epsilon,h)$ can be calculated from the general expression for the 
Gibbs free energy Eqs.~(\ref{DeltaG}) and (\ref{Fe}). The 
differentiation with respect of the "chemical potential" of Cooper 
pairs $\mu_{\rm CP}=-a_0 \epsilon,$ according to the relation 
$N_{\rm CP}=(\partial G/ \partial\mu_{\rm CP})_{T,H},$ gives 
\begin{equation} 
  n(\epsilon,h) = {N\over s} \Big\langle\left|\Psi_n\right|^2\Big\rangle 
  = \frac{1}{a_0}\frac{\partial}{\partial\epsilon}F(\epsilon,h) 
  = - \frac{\kb\tc}{a_0} S(\epsilon,h;c). 
\label{selfn} 
\end{equation} 
This formula can be alternatively derived by summation of the Rayleigh-Jeans 
asymptotics of the energy distribution of the fluctuation Cooper pairs 
\begin{equation} 
  n(\epsilon,h) = {1\over V} \sum_{{\bf p},p_z,j} 
  \frac{\kb T}{\eps_j(\textbf{p},p_z) + a} 
  = {N\over s} \Lop^{\rm (LD)}\!\!\!\int\limits_{|{\bf p}|< p_c} 
  \frac{d(\pi {p}^2)}{(2\pi\hbar)^2} 
  \frac{\kb \tc}{{p}^2 /2m_{ab}+ a_0 \epsilon}, 
\end{equation} 
see for example the monograph by Patashinskii and
Pokrovsky.\cite{Patashinskii} 

Let us give an illustration for zero magnetic field.  In this case for
the density of fluctuation Cooper pairs, using \Eref{F0} and
\Eref{LDln}, we obtain
\begin{equation} 
  n(\epsilon,0)={F_0\over a_0} \; 
  2\ln\frac{
    \sqrt{c+\epsilon} + \sqrt{c+r+\epsilon}
  }{
    \sqrt{\epsilon} + \sqrt{\epsilon+r}
  }. 
\label{LDdensity} 
\end{equation} 
This formula sets the stage for the self-consistent treatment of the
order parameter fluctuations in the LD model in which the nonlinear
term is replaced by its average. The idea has its origin in the
Maxwell consideration of the ring of Saturn; probably it is the first
work on collective phenomena in physics.  Having no possibility to
consider motion of all particles in detail we must search for some
approximation. Within a self-consistent picture, the motion of every
particle creates an average potential in which the others are
moving. From the dust of the ring of Saturn to the Cooper pairs in
cuprates the idea is the same, only the mechanics slightly changes. In
the self-consistent approximation the nonlinear term in GL equations
gives an addendum to the linear one
\begin{equation} 
  a_{\rm ren}(\epsilon,h)
  = a_0\epsilon+b\;n \left( \frac{a_{\rm ren}}{a_0},h \right), 
\end{equation} 
where the coefficient $b=\tilde b N/s$ can be expressed via the jump
of the heat capacity $\Delta C$ at the phase transition or, which is
more convenient for the high-$\tc$ cuprates, via the extrapolated to
zero temperature penetration depth $1/\lambda_{ab}^2(T)=\mu_0 n(T)
e^{* 2}/m_{ab},$ $\;n(T)=-a(T)/b,$
\begin{equation} 
  b=\frac{a_0^2}{\tc\, \Delta C}= 2\mu_0 \left( 
  \frac{\pi\hbar^2\kappa_{_{\rm GL}}}{\Phi_0 m_{ab}} \right)^2,\qquad 
  \tc\, \Delta C= \frac{1}{8\pi^2\mu_0} 
  \left(\frac{\Phi_0}{\lambda_{ab}(0)\xi_{ab}(0)}\right)^2, 
\label{bGL} 
\end{equation} 
where $\kappa_{_{\rm GL}} \equiv \lambda_{ab}(0)/\xi_{ab}(0)$ is 
the GL parameter. One can easily check that \Eref{bGL} has the same form 
in Gaussian units, where $\mu_0^{\rm (Gauss)}=4\pi.$ 
Introducing the renormalized reduced 
temperature $\epsilon_{\rm ren}>0$ for the normal phase we have 
the self-consistent equation 
\begin{equation} 
  \epsilon_{\rm ren}=\ln{T\over \tc}+ {b\over a_0} n(\epsilon_{\rm ren},h), 
\label{selfepsilon} \end{equation} 
where $n(\epsilon,h)$ is calculated by means of Gaussian saddle 
point approximation.\cite{ulah} For the LD model this equation, by virtue of 
\Eref{LDdensity}, takes the form 
\begin{equation} 
  \epsilon_{\rm ren}=\ln{T\over \tc}+ \egi \; 
  2\ln\frac{\sqrt{c+\epsilon_{\rm ren}} 
  +\sqrt{c+\epsilon_{\rm ren}+r}}{ 
  \sqrt{\epsilon_{\rm ren}}+\sqrt{\epsilon_{\rm ren}+r}} 
  =\epsilon + \egi\, \Lop^{\rm (LD)}\ln 
  {c+\epsilon_{\rm ren}\over \epsilon_{\rm ren}}, 
\label{epsh0} 
\end{equation} 
where the dimensionless parameter 
\begin{equation} 
  \egi \equiv {bF_0\over a_0^2}= 2\pi\mu_0 {N\over s} 
  \left({\lambda_{ab}(0)\over \Phi_0} \right)^2 \! \kb \tc 
  = {1\over 4\pi \xi_{ab}^2(0)}{N\over s}{\kb\over \Delta C} 
\label{GiNumber} 
\end{equation} 
is closely related to the Ginzburg number; cf. \Eref{heat} which now reads
\begin{equation} 
  {C(\epsilon)\over\Delta C}= \egi \Lop \Cop {1\over\epsilon}, 
\end{equation} 
and the review article by Varlamov~\etal\cite{varlall} 
At $\tc,$ for $\egi \ll r \ll c,$ \Eref{epsh0} gives 
\begin{equation} 
  \epsilon_{{\rm ren},\,c}\approx \egi \ln{4c\over r} 
\end{equation} 
and the effective heating $\Delta T = \tc\epsilon_{{\rm 
ren,}\,c}$ constrains the fluctuation variables at $\tc.$ 
To provide an order estimate we take for illustration 
$s_{\rm eff}=1$~nm, $\lambda_{ab}(0)= 207$~nm, $\tc=100$~K, $\xi_{ab}(0)= 
2.07$~nm, $\kappa_{_{\rm GL}}=100,$ $\kb= 1.381\times10^{-23}$~J/K. 
The substitution of these values in \Eref{GiNumber} gives 
\begin{equation} 
  \egi = {8\pi^2\times 1.381 \over1000} \approx 11\%, \qquad 
  \frac{\egi}{6\kappa_{_{\rm GL}}^2}\approx 2 \times 10^{-6}. 
\end{equation} 
In the case of nonzero magnetic field the self-consistent equation 
for the renormalized reduced temperature, \Eref{epsh0}, 
according to Eqs.~(\ref{entropy1}), (\ref{selfn}), and (\ref{selfepsilon}), 
takes the form 
\begin{equation} 
  \epsilon_{\rm ren}=\ln{T\over \tc} + \egi \, 
  \Lop \left[-\psi\!\left(\frac{\epsilon_{\rm ren}+h} {2h} 
  \right) + \psi\!\left(\frac{c+\epsilon_{\rm ren}+h} {2h} 
  \right)\right], 
\end{equation} 
or, within the LD model, 
\begin{align} 
  \epsilon_{\rm ren}=\ln{T\over \tc}+ \egi 
  \int_{0}^{\pi/2}\frac{d\phi}{\pi/2} 
  \left[-\psi\!\left(\frac{\epsilon_{\rm ren}+h+r\sin^2\phi} {2h} 
  \right)\right. \nn \\ 
  \qquad\qquad\qquad
  \left. + \, \psi\!\left(\frac{c+\epsilon_{\rm ren}+h+r\sin^2\phi} {2h} 
  \right)\right], 
\label{epsilonselfcons} 
\end{align} 
cf. also \Rref{Livanov:00}.  For weak magnetic fields, $h\ll \epsilon,$
using the asymptotic formula for the digamma function, \Eref{psiasy},
we recover \Eref{epsh0}.  The formulae pointed out could be easily
programmed for the self-consistent LD fit to the paraconductivity near
to the critical temperature $\tc.$ With the foregoing discussion we
finish the analysis of the thermodynamics of layered
superconductors. We only note that all final formulae can be used to
fit the experimental data.  Before proceeding however, for reliability
sake, it is necessary to check if the formulae implementation
correctly reproduces the 3D limit case $r\rightarrow \infty.$
 
\subsection{3D test example} 
 
\noindent 
Every layered superconductor near the critical point $|\epsilon|,h\ll
r$ displays 3D behavior. For high-$\tc$ cuprates, however, $r\ll 1$
and 3D behavior can be observed only in crystals of extremely high
quality.  Due to fluctuation of the stoichiometry and of the $\tc$ 3D
regime of Gaussian fluctuations may not occur. However there are many
conventional layered compounds with moderate anisotropy, $r \lesssim
1,$ to which the 3D behavior has broader applicability. The 3D case
can be derived as $(r\rightarrow \infty)$-asymptotics if the parabolic
band approximation $\omega_1(\theta)\approx r\theta^2/4,$
\Eref{parabolic}, is substituted into the $\Lop$ operator,
\Eref{layerLD}. Using the variable $x(\epsilon,h)$ from
Eq.~(\ref{xeh}) and a new dimensionless variable $q$, defined as
\begin{equation} 
  q       =        \sqrt{r\over 8h}{sp_z\over \hbar},\qquad 
  q^2     \equiv {r \over 8h} \theta^2,              \qquad 
  d\theta =      2 \sqrt{2h\over r} dq, 
\end{equation} 
we get for the regularized sum of logarithms in \Eref{sumln} 
the local approximation 
\begin{equation} 
  \Lop^{\rm (LD)} \Reg{\zeta} \sum_{n=0}^{\infty} 
  \ln\!\left(n+{1\over2}+{\epsilon\over2h}\right)\approx 
  2\sqrt{2h\over r}\; 
  \Reg{\zeta}\sum_{n=0}^{\infty} \int_{-\infty}^{+\infty}\frac{dq}{2\pi} 
  \ln\!\left(n+x+q^2\right). 
\end{equation} 
The UV regularization in this expression is carried out with the help of 
the equation 
\begin{equation} 
  \Reg{\zeta}\sum_{n=0}^{\infty} \int_{-\infty}^{+\infty}\frac{dq}{2\pi} 
  \ln\!\left( n + x + q^2 \right) 
  = \zeta\!\left( -{1\over2},x \right), 
\label{Todor89} 
\end{equation} 
which can be easily proved using derivatives of $\zeta$-functions 
\begin{equation} 
  \frac{d}{dx}\zeta(\nu,x) = - \nu \zeta( \nu + 1, x ). 
\end{equation} 
The second derivative of \Eref{Todor89} is trivially convergent; the
essence of the $\zeta$-function regularization lies in the omission of
an arbitrary linear function $A(c) x + B(c)$, being analytical with
respect to $\epsilon$ and therefore irrelevant to the critical
behavior, cf. \Eref{LinUV}.  In fact $c \simeq 1$ but having dropped
$A(c)$ and $B(c)$ we can obtain the local approximation, $|\epsilon|,
h \ll c,$ as $c\rightarrow \infty$ even if $A(\infty)=\infty$ and
$B(\infty)=\infty.$ The substitution of this UV regularization in
\Eref{sumln}, using \Eref{LDparameter}, gives the result by
Mishonov\cite{todor} for the fluctuation part of the Gibbs free energy
\begin{equation} 
  G(T,H) = V F(\epsilon,h)
         = \frac{\sqrt 2}{2\pi}\,\kb T \, 
           \frac{V}{\xi_a(0)\xi_b(0)\xi_c(0)}\, h^{3/2} \, 
           \zeta\!\left( -{1\over 2},{1\over 2} + {\epsilon\over 2h} \right). 
\label{zeta1989} 
\end{equation} 
This result was confirmed by Baraduc~\etal,\cite{Baraduc} using the
same notations, with the $\zeta$-function presented implicitly.
 
In order to bridge the 3D result with the notations introduced for layered 
systems we can rewrite the coefficient in \Eref{zeta1989} as 
\begin{equation} 
  \frac{\sqrt 2}{2\pi} \, \frac{\kb T}{\xi_a(0)\xi_b(0)\xi_c(0)} 
  = 4\sqrt{2\over r}\left( {1\over 2}M_0B_{c2}(0) \right). 
\end{equation} 
Now differentiation $F(\epsilon,h)$ with respect to the magnetic field 
we obtain for the dimensionless magnetization, in 
agreement with the result by Kurkij\"arvi, Ambegaokar and 
Eilenberger\cite{kurk} 
\begin{equation} 
  {\tilde m}(\epsilon,h) = 3\left( {2 \over r} \right)^{1/2} \! \sqrt{h} 
  \left[ 
    \zeta\!\left(
      -{1\over 2},{1\over 2}+{\epsilon\over 2h}
    \right) - {1\over 3} 
    \zeta\!\left(
      {1\over 2},{1\over 2}+{\epsilon\over 2h}
    \right) \frac{\epsilon}{2h} 
  \right]. 
\label{zetaM} 
\end{equation} 
The subsequent differentiation with respect to the magnetic field
gives the differential susceptibility. In the particular case of
strong magnetic fields, $\epsilon \ll h,$ the local approximation to
the GL model, \Eref{zetaM}, gives the well known-result by
Prange\cite{prange} with an anisotropy correction
multiplier\cite{todor} $\xi_{ab}(0)/\xi_{c}(0)$
\begin{equation} 
  \tilde{m}(0,h) = 3 \sqrt{2} \times 0.0608885 \; \sqrt{h\over r}, \quad 
  -M = 3 \pi^{1/2} \zeta\!\left(-{1\over 2},{1\over 2}\right) 
      \frac{\kb\tc}{\Phi_0^{3/2}}\frac{\xi_{ab}(0)}{\xi_{c}(0)}\sqrt{\mu_0 H}, 
\label{Prange} 
\end{equation} 
where for the values of the $\zeta$-function we have 
\begin{align} 
  \zeta\!\left(-{1\over 2},{1\over 2}\right) 
  & = \left[-1+{1\over\sqrt 2}\right] \zeta\!\left(-{1\over 2}\right) 
    = \texttt{Zeta[-1/2,1/2]} 
    = 0.0608885\dots,  \nn \\ 
  \zeta\!\left(-{1\over 2}\right) &= \texttt{Zeta[-1/2]} = -0.207886\dots \;. 
\label{608} 
\end{align} 
The syntax \texttt{Zeta[\ldots]} is used in the commercial software
\Mathematica.\cite{Wolfram} We stress, however, that these are only
test mathematical asymptotics for $c\rightarrow \infty$. For the
magnetization, as well as for every quantity exhibiting UV divergences
in the local limit, the nonlocal effects are strongly pronounced
simply because the contribution of high momenta is significant. That
is why the local approximation could be quantitatively fairly good for
fitting to the data for fluctuation conductivity and heat capacity.
For the magnetization in strong magnetic field regime we have to take
into account the effect of nonlocality by fitting the energy cutoff
parameter $\ecut$.  A systematic procedure for determination of the
parameters of the GL theory is developed in the next section.

\section{Some remarks on the fitting of the GL parameters} 
 
\subsection{Determination of the cutoff energy $\ecut$} 
 
\noindent 
Let we start with designing a general procedure to fit some 
parameters of the GL theory which employs only data for the in-plane 
paraconductivity. Later on we shall address  the advantage 
of investigating  several variables simultaneously. The first 
step is to extract the fluctuation part of the conductivity from 
the temperature dependence of the resistivity $R(T)$. For layered 
cuprates the resistivity of the normal phase is to within good accuracy a 
linear function of temperature, $R_N(T)=A_R+B_R T,$ and we can 
fit the coefficients $A_R$ and $B_R$ far enough from the critical 
temperature $\tc,$ \eg in the temperature interval  $(1.5\,\tc,3\,\tc).$ 
After that we can determine the experimental data for 
the fluctuation conductivity 
\begin{equation} 
  \sigma_i = R(T_ i)^{-1} - (A_R + B_R T_i)^{-1} 
\end{equation} 
for all experimental points $i = 1,\dots, N_{\rm exp}.$ For bi-layered
cuprates, such as YBa$_2$Cu$_3$O$_{7-\delta}$ and
Bi$_2$Sr$_2$CaCu$_2$O$_8$, one can attempt fitting the data with the
formula for the bilayered model, \Eref{MT1}, where an arbitrary
life-time $\tilde\tau_{\rm rel}$ and cutoff parameter $c$ are included
in the interpolation
\begin{equation} 
  \sigma(\epsilon;\tilde\tau_{\rm rel},r,w,c) = \frac{e^2}{16\hbar}
  \tilde\tau_{\rm rel}{N\over s} \left[f_{\rm MT}(\epsilon;r,w) 
  - f_{\rm MT}(c+\epsilon;r,w) \right], 
\end{equation} 
where 
\begin{equation} 
  f_{\rm MT}(\epsilon;r,w)\equiv \frac{\epsilon+{1\over 2}rw} 
  {\sqrt{\left(\epsilon^2+rw\epsilon\right) 
  \left(\epsilon^2+rw\epsilon+{1\over 4}r^2w\right)}} 
  = \Lop^{\rm (MT)} f_{\rm APS}(\epsilon,h=0) 
  = \Lop^{\rm (MT)} {1\over \epsilon}. 
\end{equation} 
As a next step, if necessary, one may fit the data using logarithmic
plot that generates the dimensionless deviations from the
$\ln(\sigma_{ab}(T))$-values,
\begin{equation} 
  x_i(r,w,c) = 
  \ln\left[ 
    f_{\rm MT}(\epsilon_i;r,w) - f_{\rm MT}(c + \epsilon_i; r,w) 
  \right] 
  - \ln\!\left(
    \frac{
      \sigma_i
    }{
      {e^2 \over 16\hbar}{N \over s}
    }
    \right). 
\end{equation} 
For the $x_i$ data we can calculate the mean value, the averaged square 
\begin{equation} 
  \langle x  \rangle = {1\over N_{\rm exp}}\sum_{i=1}^{N_{\rm exp}} x_i,\qquad 
  \langle x^2\rangle = {1\over N_{\rm exp}}\sum_{i=1}^{N_{\rm exp}} x_i^2, 
\end{equation} 
and the dispersion 
\begin{equation} 
  S(r,w,c) = \langle x^2\rangle - \langle x \rangle^2. 
\end{equation} 
The fitting procedure is then reduced to numerically finding the
minimum of the dispersion
\begin{equation} 
  S(r_0,w_0,c_0) \le S(r,w,c) 
\end{equation} 
in the space of parameters $(r,w,c).$ We have to start from some
acceptable set of parameters, for example, $r_0={1\over7},$ $ w_0=1,$
and $c_0={1\over2}$, and to search for the minimal value in certain
range, \eg $ r_0=\in(0,\;1),$ $w_0\in(1,30)$ and
$c_0={1\over2}\in(0.2,2).$ It is possible that the parameters of the
normal resistivity be corrected by the same procedure for minimization
of the dispersion $S(r,w,c,A_R,B_R)$. The contemporary methods of the
mathematical statistics, such as the bootstrap and Jack-knife, can
tell us how reliable is the set of the fitted parameters; the simplest
possible realization is to decrement sequentially $N_{\rm exp}$ by one
and to investigate the distribution of the fitted GL parameters at
every step. For example, the $w$ parameter is almost inaccessible
since for $w=1$ and $w \rightarrow \infty$ we have LD-type temperature
dependence of the paraconductivity. On the other hand if we try to fit
the paraconductivity far from the critical temperature, \eg $T\in
(1.02\,\tc, 1.15\,\tc)$ we can easily find some estimate for the
cutoff parameter $c.$ In any case a good fit would be useful because
as a by-product we determine the life-time of the fluctuation Cooper
pairs
\begin{equation} 
  \tilde\tau_{\rm rel} = \exp\left(-\langle x\rangle (r_0,w_0,c_0)\right). 
\end{equation} 
The same procedure can be applied to the magnetic susceptibility at
vanishing magnetic field which, according to \Eref{tauratio}, is
proportional to the conductivity, or for the susceptibility in the LD
model which, according to Eqs.~(\ref{chiLD}), (\ref{cutLD}), and
(\ref{GiNumber}), reads
\begin{equation} 
  -\chi_{_{\rm LD}}(\epsilon)
  = {1\over 6} {\egi \over \kappa^2_{_{\rm GL}}}
    \left[ 
      {1 \over \sqrt{\epsilon(\epsilon+r)}}
     -{1 \over \sqrt{(c+\epsilon)(c+\epsilon+r)}} 
    \right], 
\end{equation} 
where 
\begin{equation} 
  {\egi \over \kappa^2_{_{\rm GL}}} 
  = 2\pi\mu_0 \frac{\kb\tc}{\Phi_0^2}\xi_{ab}^2(0){N\over s} 
  = \frac{M_0}{H_{c2}(0)}. 
\end{equation} 
The general formula for the conductivity, Eqs.~(\ref{Lsigma}), (\ref{Csigma}), 
\begin{equation} 
  \sigma_{ab}(\epsilon,h) = {\pi\over8}\frac{\tau_{\rm rel}} {R_{\rm QHE}} \;
                            \Lop^{\rm (LD)} \Cop f_{\rm APS}(\epsilon,h), 
\end{equation} 
which for single layered superconductor reads, cf. \Eref{LCM}, 
\begin{align} 
  \sigma_{ab}(\epsilon,h;r,C)  = &\, \tilde\tau_{\rm rel} 
  \frac{e^2}{16\hbar s_{\rm eff}} \; {2\over h^2} 
  \int_0^{\pi/2}\frac{d\phi}{\pi/2}\left\{ \left(\epsilon+r\sin^2\phi\right) 
  \left[\psi\!\left({1\over2} + \frac{\epsilon+r\sin^2\phi}{2h}\right) 
                          \right. \right. \nn \\ 
& \left.-\psi\left(1+\frac{\epsilon+r\sin^2\phi}{2h}\right) 
  + \frac{h}{\epsilon+r\sin^2\phi}\right] \nn \\ 
& - \left(c+\epsilon+r\sin^2\phi\right) 
  \left[
   \psi\!\left({1\over2}+\frac{c+\epsilon+r\sin^2\phi}{2h}\right)
                                  \right. \nn \\ 
& \left.\left. 
  - \psi\!\left(1+\frac{c+\epsilon+r\sin^2\phi}{2h}\right) 
  + \frac{h}{c+\epsilon+r\sin^2\phi}\right] \right\} 
\label{LCsigma} 
\end{align} 
gives another possibility for determining the energy cutoff parameter
$c.$ As most appropriate regime we recommend that the measurements of
the conductivity as a function of the magnetic field to be carried out
at the critical temperature $T=\tc.$ In this case for strong magnetic
field, $h\gg r,$ the layered superconductors with strong anisotropy
$r\ll 1$ show 2D behavior. The substitution $\epsilon =0$ in
\Eref{LCsigma} gives another universal law derived within the GL
theory with energy cutoff
\begin{align} 
  {1\over2}\frac{B}{B_{c2}(0)} 
  \frac{
    \sigma_{ab}(\epsilon=0,h)
  }{ 
    \tilde\tau_{\rm rel} (e^2 / 16\hbar s_{\rm eff})
  } 
   & = \Cop\, {h\over2}\,f_{\rm APS}(\epsilon=0,h) \nn \\ 
   & = \frac{\pi\hbar}{e} s_{\rm eff} 
       \frac{\xi^2_{ab}(0)}{\tau_0} \; B \sigma_{ab}(\tc,B) 
     = U_{\sigma}\!\left({2\over c} {B\over B_{c2}(0)}\right), 
\end{align} 
where $y=2h/c,$ cf. \Eref{UM}, 
\begin{equation} 
  U_{\sigma}(y) = {2\over y}\left[\psi\!\left(1+{1\over y}\right) 
                  -\psi\!\left({1\over2}+{1\over y}\right)\right], 
\label{Usigma} 
\end{equation} 
$U_{\sigma}(0)=1,$ and $U_{\sigma}(\infty)=0.$ At best, the 
universal dimensionless conductivity $U_\sigma\propto B\sigma(B)$ 
and magnetization $U_M \propto M$ have to be fitted simultaneously 
using the data for the same crystal and common dimensionless 
argument $\propto B.$ Similar universal scaling law for the heat 
capacity can be derived from \Eref{Ccut} 
\begin{equation} 
  \frac{2}{\zeta\left(2,{1\over2}\right)}\frac{B}{B_{c2}(0)} 
  \frac{C(\tc, B)}{\egi\Delta C}=U_C\left({2\over c} 
  {B \over B_{c2}(0)}\right), 
\end{equation} 
where 
\begin{equation} 
  U_C(y) = 1 -\frac{\zeta\!\left(2,{1\over2}+ 
          {1 \over y}\right)}{\zeta\!\left(2,{1\over2}\right)}, 
\label{UniC} 
\end{equation} 
but the accuracy of thermal measurements is probably not high 
enough in order for this to be experimentally confirmed.
\pagebreak[3] 
 
\subsection{Determination of the coherence length $\xi_{ab}(0)$} 
 
\noindent 
The fit of every fluctuation variable as a function of the dimensional
magnetic field $h,$ the conductivity 
\begin{equation} 
  \sigma(\epsilon,h) = \sigma(\epsilon) + \Delta\sigma(\epsilon,h), 
\end{equation} 
for example, provides a method for determination of $B_{c2}(0)$ and
$\xi_{ab}(0).$ At weak magnetic fields, $h\ll \epsilon$, the
magnetoconductivity is proportional to the square of the magnetic
field $\Delta\sigma(\epsilon,h)\equiv
\sigma(\epsilon,h)-\sigma(\epsilon)\propto B^2.$ For this small
negative quantity, $0<-\Delta\sigma(\epsilon,h)\ll \sigma(\epsilon),$
the APS result, \Eref{APSlowh}, reads\cite{EPL}
\begin{equation} 
  -\Delta\sigma(\epsilon,h)\approx {h^2\over4} \frac{\partial^2}{\partial 
   \epsilon^2}\sigma(\epsilon), 
\label{magnetocon} 
\end{equation} 
where $ h=2\pi\xi_{ab}^2(0)B_z/\Phi_0.$ The common multiplier $\tau_0$
from \Eref{APSsigma} is obviously canceled in this relation because,
roughly speaking, the transport takes time even in the presence of
magnetic field.  We note that a multiplier $\tilde\tau_{\rm rel}$ was
misintroduced by M.~V.~Ramallo in \Rref{EPL} in the right-hand-side of
the above equation (see Eq.~(4) in \Rref{EPL}). Thereby the old
experimental data in \Rref{EPL} have been apparently processed by
employing erroneous expression and therefore the discussion related to
Fig.~2 in \Rref{EPL} is physically unsound. As a consequence, the
life-time constant of metastable Cooper pairs in cuprates is still
waiting for its first experimental determination.  Nevertheless the
novel theoretical result that the life-time constant $\tau _0$ and the
diffusion coefficient of the fluctuation Cooper pairs
$\xi_{ab}^2(0)/\tau_0$ can be determined from the $\sigma/\chi$-ratio
remains unchanged.
 
Returning to \Eref{magnetocon} we note that 
after two-fold integration of the relation (\ref{magnetocon}) 
in some temperature interval, \eg $(\epsilon_a,\;\epsilon_b)=(0.03, 0.09),$ 
the ``noise'' in the experimental data is already irrelevant and we can rewrite 
\Eref{magnetocon} as\cite{EPL} 
\begin{equation} 
  \xi_{ab}(0)=l_B 
  \left[ 
    \frac{\displaystyle\int_{\epsilon_a}^{\epsilon_b}{d}\epsilon' 
      \int_{\epsilon'}^{\epsilon_b} \left(-\Delta 
      \sigma(\epsilon'',h)\right) d\epsilon''
    }{\displaystyle \sigma(\epsilon_a) - \sigma(\epsilon_b) 
      +(\epsilon_b-\epsilon_a)\frac{d\sigma}{d\epsilon}(\epsilon_b)
    } 
  \right]^{1/4} 
\label{length}, 
\end{equation} 
where $l_B$ is the magnetic length 
\begin{equation} 
  l_B = \sqrt{\Phi_0\over\pi B} 
      = \sqrt{\hbar\over e B} 
      = \frac{25.6\; {\rm nm}}{\sqrt{B(T)}}. 
\end{equation} 
 
For practical application we have to take into account that far from
the critical temperature, even for $T-\tc = 15\%\; \tc$ the
fluctuation conductivity is negligible
$\sigma(0.15)\approx\sigma(\infty)=0.$ That is why in acceptable
approximation we can take $\epsilon_b=0.15\approx "\infty".$ For
$\epsilon>\epsilon_b$ the temperature dependence of the
magnetoconductivity in the numerator of \Eref{length} can be an
extrapolated LD fit.
 
However, due to the strong critical behavior $-\Delta \sigma \propto
h^2/\epsilon^3$ for $\epsilon \gg r$ the influence of the interval
$(\epsilon_b,\infty)$ can be neglected. In such a way, after a partial
integration, we arrive at a simpler equation for determination of the
in-plane coherence length, cf. \Rref{EPL},
\begin{equation} 
  \xi_{ab}(0) \approx l_B \left[\frac{1}{\sigma(\epsilon)} 
  \left( 
    \int_{\epsilon}^{\infty}\epsilon' 
    \left(-\Delta \sigma(\epsilon^{\prime},h)\right) d\epsilon^{\prime} 
    - \epsilon \int_{\epsilon}^{\infty} 
    \left(-\Delta \sigma(\epsilon'',h)\right) d\epsilon'' 
  \right) 
  \right]^{1/4}= \text{const}, 
\label{best} 
\end{equation} 
where the integrations should be performed in the whole experimentally
accessible temperature range above $(1+\epsilon)\tc.$ This result of
the Gaussian fluctuation theory does not depend upon the $\tau_0$
parameter, effective mass of Cooper pairs $m_{ab},$ and the space
dimensionality. We consider this procedure for determination of the
coherence length $\xi_{ab}(0)$ as being the best one, as it is
model-free and does not depend on the multilaminarity of the
superconductor, \ie on the dispersion of Cooper pairs in $c$-direction
$\varepsilon_{c,j}(p_z)$. Equation (\ref{best}) has the same form for
both strongly anisotropic high-$\tc$ cuprates and bulk conventional
dirty alloys. Of course, methods particularly based on the proximity
to the critical line $H_{c2}(T)$ can be very useful in determining
$\xi_{ab}(0)$ especially in the case of strong magnetic fields.  For
example, \Eref{APSlowh} gives another appropriate formula
\begin{equation} 
  \sigma_{ab}(\epsilon,h) \approx \tilde \tau_{\rm rel} 
  \frac{e^2}{16\hbar}{N\over s} 
  \frac{4}{\sqrt{(\epsilon+h)(\epsilon + h +r)}} 
\label{sigmaHc2} 
\end{equation} 
applicable for $\egi \ll \epsilon+h\ll h.$ Similar result, cf. also
\Eref{localmagn}, 
\begin{equation} 
  M= -M_0 \; \tilde m \approx - \frac{\kb \tc}{\Phi_0}{N\over s} 
  \frac{h}{\sqrt{(\epsilon+h)(\epsilon+h+r)}} 
\label{MHc2} 
\end{equation} 
can be derived under the  same physical conditions from the formula 
for the fluctuation magnetic moment, \Eref{LCM}, using the 
approximations for $0<x\ll 1,$ 
\begin{equation} 
  \Gamma(x)\approx -\ln x -{1\over2} \ln(2\pi),\qquad 
  \psi(x) \approx - {1\over x}. 
\end{equation} 
The experimental investigation of the conductivity, \Eref{sigmaHc2},
and magnetization, \Eref{MHc2}, is probably the best way to extract
the upper critical field $H_{c2}(T)$ for high-$\tc$ cuprates; the
$H\sigma/M$ quotient near the critical line is $2/3$ of the
$\sigma/\chi$ quotient for weak magnetic fields.

\subsection{Determination of the Cooper pair life-time constant $\tau_0$} 
 
\noindent 
Having a reliable estimate for the coherence length, the life-time
constant of the metastable Cooper pairs above $\tc$ can be determined
via the $\sigma/\chi$-quotient, \Eref{tauratio}. We believe that
this method will become a standard procedure in the physics of
high-$\tc$ materials.  Certainly the most transparent method is just
the fit to the phase angle of high-frequency complex fluctuation
conductivity
\begin{equation} 
  \phi_\sigma(\omega \tau(\epsilon_{\rm ren}))
  = \arctan \frac {\sigma''(\omega)} {\sigma'(\omega)} 
  = \arctan \frac{
      \Lop\, \varsigma_2\!\left(\omega\tau_0/\epsilon_{\rm ren}\right)
    }{
      \Lop\, \varsigma_1\!(\omega\tau_0/\epsilon_{\rm ren})
    }. 
\label{phisigma} 
\end{equation} 
The state-of-the-art electronics gives such a possibility, but
unfortunately the first experiments of the type\cite{Boot,Anlage} was
not performed in the Gaussian region. For the development of Gaussian
spectroscopy which will give results relevant for the microscopic
mechanisms of superconductivity we recommend the use of the
conventional thin films and high-quality low temperature cuprate
films, such as Bi$_2$Sr$_2$CaCu$_2$O$_8.$
 
\subsection{Determination of the Ginzburg number and penetration 
depth $\lambda_{ab}(0)$} 
 
\noindent 
The applicability of the self-consistent approximation in the theory
of fluctuation phenomena in superconductors is strongly limited by the
quality of the samples. The fluctuation of the critical temperature
$\Delta \tc$, \eg due to the oxygen stoichiometry in cuprates, should
be small enough, $\Delta \tc \ll \egi\tc,$ and this has to be verified
empirically.  If the $\sigma/\chi$ ratio remains temperature
independent for $\epsilon < 3\%$ and both $\sigma(\epsilon)$ and
$\chi(\epsilon)$ demonstrate weak deviation from the LD fit obtained
from the range $\epsilon\in (3\%, 9\%),$ this could be considered as a
hint in favor of the self-consistent approximation. In this case
$\egi$ can be fitted by substituting the solution $\epsilon_{\rm
ren}(\epsilon)$ of \Eref{epsh0} into the LD fit to $\sigma^{\rm
(LD)}(\epsilon_{\rm ren})$ and $\chi^{\rm (LD)}(\epsilon_{\rm ren}).$
We note that the reliability in fitting $\egi$ is determined by the
condition whether the self-consistent approach and the use of
$\epsilon_{\rm ren}$ significantly improve the accuracy of the fit to
the experimental data near $\tc.$ According to \Eref{GiNumber} we can
parameterize $\egi$ with the help of the penetration depth
$\lambda_{ab}(0).$ In any case, an evaluation of such type should be a
part of the complete set of GL parameters of the superconductor.
Another possibility for the thermodynamic determination of the
penetration depth $\lambda_{ab}(0)$ is provided by the jump in the
specific heat at the critical temperature, \Eref{bGL}.  As a rule the
accuracy of the determination of the penetration depth by the
thermodynamic methods cannot be high, especially for high-$\tc$
cuprates where the phonon part strongly dominates.  An acceptable
value of $\ln\kappa_{_{\rm GL}}$ derived from the heat capacity is
necessary for the establishment of a coherent understanding of the
superconductivity; there is no doubt that the direct investigation of
the vortex phase of the superconductors or vortex-free high-frequency
measurements constitute the best methods for determination of
$\lambda_{ab}(0).$

\section{Discussion and conclusions} 
 
\noindent 
In the attempts to systematize the available results we had to derive
in parallel new ones too. We shall summarize the most important of
them starting with remarks concerning the theory. As the ultimate
result we consider the representation of the fluctuation part of the
Gibbs free energy by the Euler $\Gamma$-function \Eref{theresult} in
Gaussian approximation. This result trivializes the derivation of all
thermodynamic variables, such as fluctuation magnetization \Eref{LCM},
or fluctuation heat capacity \Eref{tedious}. To our knowledge this is
a novel result, but we find it strange that it remained unobserved
given the great attention which the fluctuations in high-$\tc$
superconductors have attracted.  The importance of fluctuations was
mentioned even in the classical work by Bednorz and
M\"uller. Fluctuations in superconductors were among the main topics
in many scientific activities; the $\Gamma$-function is well-known to
all physicist; the mathematical physics behind the 2D statistical
mechanics is well developed, polygamma functions can be found in a
number of BCS papers, and finally the solution turns out to be on a
textbook level.  Just the same is the situation for the 3D GL
model. In this case the solution for the free energy is given in terms
of the Hurwitz $\zeta$-functions. Analogous result gave the name of
one of the most powerful methods in the field
theory --- $\zeta$-function method for ultraviolet regularization, but
this method was never applied to the most simple problem of a 3D GL
model related to numerous experiments in the physics of
superconductivity.
 
Another simple but useful detail is the layering operator $\Lop$, \Eref{layer}, which 
allows us to extend the 2D result onto layered superconductors and even to 3D 
superconductors. The method can be applied not only to the 
thermodynamic variables but to  the fluctuation part of the kinetic 
coefficients as well. In this way we obtained useful formulae 
for the in-plane fluctuation conductivity in perpendicular magnetic field, 
\Eref{LCsigma}, and for the high-frequency Aslamazov-Larkin conductivity in 
layered superconductors, \Eref{sigmaomegaextended}. We proposed further convenient 
$r$-$w$ parameters, \Eref{MT1}, for the bi-layered model which could be 
utilized for experimental data processing of the fluctuation phenomena 
in bi-layered cuprates, such as YBa$_2$Cu$_3$O$_{7-\delta}$ for 
example. 
 
The representation of the thermodynamic variables via polygamma
functions is very helpful at strong magnetic fields but due to the
presence of the magnetic field in denominator these results cannot be
directly applied to zero-magnetic-field limit. For small magnetic
fields, on the other hand, we have to use asymptotic formulae for
polygamma and $\zeta$-functions with large arguments. This is the
reason why the weak-magnetic-field expansion of the magnetization and
the other thermodynamic variables has so bad convergence. In order to
fit the experimental data for the magnetization in weak magnetic field
using the new analytical result for the LD model, \Eref{MlowB}, we
arrive at the problem for summation of divergent asymptotic series. At
least for experimentalists this is a nontrivial problem which led us
to give a prescription for usage of series from the theoretical
papers.  There is no doubt that the $\eps$-method is one of the
brilliant achievements of the applied mathematics of XX
century. However, it turns out that this method was not cast in an
suitable form to be employed by users like experimentalists having no
time to understand how the underlying mathematics can be derived. That
is why we presented an oversimplified version of this algorithm
illustrated by a simple \textsc{fortran90} program. The latter can be
also used for calculation of the differential nonlinear susceptibility
at finite magnetic field, \Eref{kappasum}, which is another novel
result in the present work.
 
Let us now address the simple final formulae that can be directly used
for experimental data processing. First of all we advocate that the
relation between fluctuation conductivity and magnetoconductivity,
\Eref{best}, provides the best method (shortly announced in
\Rref{EPL}) for determination of the in-plane coherence length
$\xi_{ab}(0)$ in layered high-$\tc$ cuprates and conventional
superconductor superlattices and thin films. Having such a reliable
method for determination of $\xi_{ab}(0),$ the Cooper pair life-time
spectroscopy can be created\cite{EPL} on the basis of determination of
the life-time constant $\tau_0$ by the $\sigma/\chi$ quotient,
\Eref{tauratio}.
 
Usually science starts with some simplicity, thus it is surprising
that the temperature independence of the $\sigma/\chi,$ $\chi/C,$ 
and $\sigma/C$ quotients has not attracted any attention in physics. 
The question of whether the
high-$\tc$ cuprates are BCS superconductors, or they have a non BCS
behavior, consumed more paper and brought more information pollution
than that about the sense of life, about the smile of Mona Lisa.  Now
we possess a perfect tool to check whether this sacramental ${\pi\over
8}$ BCS ratio, \Eref{pi8}, still exists in the physics of high-$\tc$
superconductivity. A careful study of the relative life-time constant
$\tau_0$ by the $\sigma/\chi$ ratio, \Eref{taurel}, will provide a
unique information on the presence of depairing impurities in the
superconducting cuprates. The doping dependence of this ratio will
give important information for the limits of applicability of the
self-consistent BCS approximation. In principle the same life-time
spectroscopy can be applied to heavy fermions and other exotic
superconductors.
 
The methods we proposed in this review can be initially tested by
means of alternative methods for determination of $\xi_{ab}(0)$, \eg
from the slope of the upper critical field $H_{c2}(T)$ defined by the
fluctuation magnetization of the normal phase near the critical line,
\Eref{MHc2}, being another new result derived here, or from the
fluctuation conductivity of the LD model, \Eref{sigmaHc2}.  Addressing
the conductivity we consider that the fitting to high frequency
experimental data with the help of formulae (\ref{sigmaomega}) and
(\ref{phisigma}) will give a direct method for determination of the
relaxation time of the superconducting order parameter.  A good
monocrystal of layered cuprate or high-quality thin film with as low
as possible critical temperature could ensure the overlap of both the
suggested methods for Cooper pair life-time spectroscopy. At present
we only know\cite{EPL} that for 93~K YBa$_2$Cu$_3$O$_{7-\delta}$
$\tau_{0,\Psi}=2\tau_0= 32\;{\rm fs}.$ We hope, however, that several
experimental methods for determination of $\xi_{ab}(0)$ and $\tau_0$
will be mutually verified in the nearest future. Thus, the
investigation of the Gaussian fluctuations may become a routine
procedure in the materials science of superconductors.
 
We also believe that the development of the Gaussian spectroscopy will
lead to determination of the Ginzburg number $\egi,$ the energy
cutoff, \ie the maximal kinetic energy of the Cooper pairs $\ecut = c
\hbar^2/2m_{ab}\xi^2_{ab}(0).$ Up to now these parameters of the GL
theory are inaccessible. We hope that our derived self-consistent
equation for the reduced temperature, \Eref{epsilonselfcons}, will
stimulate experimentalists to reexamine the data for high-quality
crystals in the region close to $\epsilon+h \simeq 3\%$ in order to
extract $\egi.$
\begin{figure}[h] 
\begin{center} 
  \epsfig{file=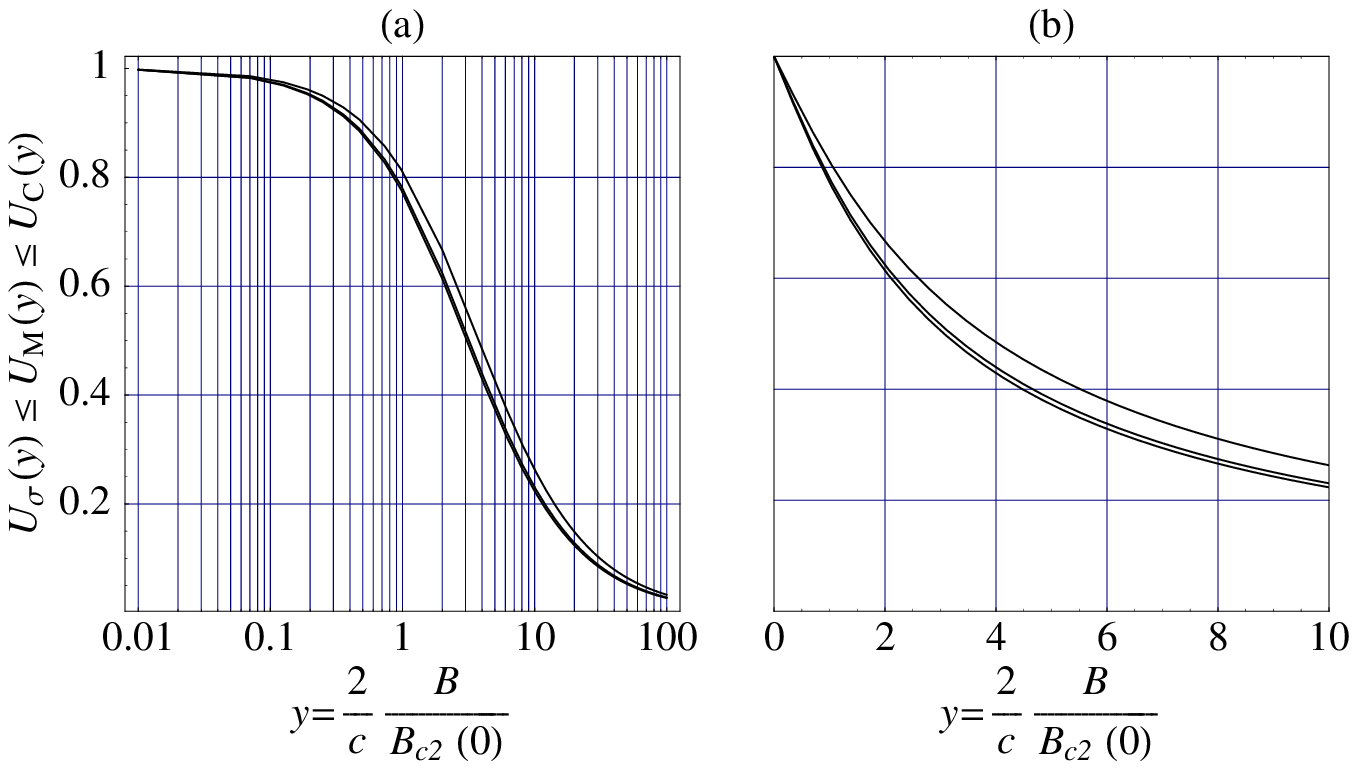,width=0.8\textwidth} 
\end{center} 
\vspace*{13pt} 
\fcaption{%
Universal scaling curves ((a) semi-logarithmic plot; (b) linear plot) 
of a quasi-2D superconductor for the fluctuation conductivity $B\sigma(\tc, 
B)\propto U_{\sigma},$ magnetic moment $M(\tc,B)\propto U_{M},$ 
and heat capacity $B C(\tc,B) \propto U_C$ versus 
dimensionless magnetic field $y=2h/c.$ The fit of the scale in 
horizontal direction gives the GL cutoff parameter $c.$ The 
scales in vertical directions are related correspondingly to 
diffusion constant of Cooper pairs $\xi^2_{ab}(0)/\tau_0$, 
effective inter-layer distance $s_{\rm eff}$, and 2D Ginzburg 
number $\egi$, cf. Eqs.~\protect{(\ref{Usigma}), 
(\ref{UM}), and (\ref{UniC})}. 
} 
\label{fig:curves} 
\end{figure} 
Virtually all final results are presented by taking into account
the energy cutoff parameter $c.$ The nonlocality corrections can be
extracted from almost all fluctuation variables, if $\epsilon + h
>10$\%, but we suggest special new experiments to be conducted for
investigation of nonlocality effects in quasi 2D superconductors at
$\tc.$ Analogous investigations for fluctuation diamagnetism for
classical bulk superconductors are already classics in physics of
superconductivity; see for example Fig.~8.5 in the well-known textbook
by Tinkham.\cite{Tinkham} The universal scaling law for the heat
capacity, \Eref{UniC}, for the magnetization, \Eref{UM}, and
conductivity, \Eref{Usigma}, versus the reduced magnetic field
$y=2h/c$ are depicted in Fig.~\ref{fig:curves}.
 
At least for conductivity the experimental confirmation for the
quasi-2D superconductors ($h\gg r,\egi$) can be easily achieved. How
different are the animals\ldots? The biochemists considered that what
is true for \textit{Escherichia coli} holds true for the
elephant. Analogously, we consider that $B\sigma(\tc,B)$ versus $B$
will be within 20\% accuracy the same for conventional Pb layers and
for strongly anisotropic underdoped Bi$_2$Sr$_2$CaCu$_2$O$_8$ in spite
of the bunch of sophisticated theories of high-$\tc$
superconductivity. The GL theory gives the scaling law, the notions
and notations, and in this sense the language for analysis of the
fluctuation phenomena. The precisely measured deviations from the GL
scaling low could give the basis for further microscopic consideration
using the methods of the statistical mechanics. This is the last
example how the development of the Gaussian fluctuation spectroscopy
could be of importance not only for the materials science but for the
fundamental physics of superconductivity as well.
 
\nonumsection{Acknowledgments} 
\noindent 
Some parts of this review are based on the lectures on statistical
mechanics, solid state theory, numerical methods and physics of
superconductivity read by one of the authors (T.M.) at the University
of Sofia. He would like to thank to many of his former students for
the help and collaboration in the early stages of this work,
especially to D.~Damianov for the collaboration on the Boltzmann
equation for the fluctuation Cooper pairs\cite{Damianov} and to
N.~Zahariev for implementing the $\eps$-algorithm in C.
It is a pleasure to acknowledge the cooperation of C.~Carballeira in
implementing the $\eps$-algorithm in \Mathematica\ and checking the
numerical equivalence for $r=0$ of Eqs.~(\ref{MlowB}) and (\ref{LCM})
using this software.  The same author appreciates the discussions with
Prof.~Vidal on the experiments concerning $\sigma/\chi$ and other
quotients.
The completion of this review would be impossible without the warm
hospitality of Prof.~Indekeu, his support and interest to this work.
 
The early stages of this study were supported by the Bulgarian NSF
($\Phi$627/1996) and the Visiting Professor Fellowship from the
Spanish Ministry of Education.  This paper was supported by the
Belgian DWTC, the Flemish Government Programme VIS/97/01, the IUAP and
the GOA.
\pagebreak[3] 

\appendix{} 
\label{appendix} 
{ 
\renewcommand{\baselinestretch}{0.8} 
\footnotesize 
\begin{verbatim} 
!+ Test driver program for subroutine Limes 
PROGRAM Test 
 
IMPLICIT NONE 
 
  INTEGER,   PARAMETER :: pr = SELECTED_REAL_KIND (30,150) 
  REAL (pr), PARAMETER :: zero = 0.0, one = 1.0 
  REAL (pr)            :: S(0:137), C(0:137), x, xi, arg 
  REAL (pr)            :: rLimes 
  REAL (pr)            :: err 
  INTEGER              :: N 
  INTEGER              :: i 
  INTEGER              :: i_Pade 
  INTEGER              :: k_Pade 
  INTEGER              :: is 
 
  WRITE (*, '(15X,A)') '+---------------------------------------------+' 
  WRITE (*, '(15X,A)') '|   Test driver program for subroutine Limes  |' 
  WRITE (*, '(15X,A)') '|               Calculate Ln[x]               |' 
  WRITE (*, '(15X,A)') '+---------------------------------------------+' 
  WRITE (*, '(A)') ' ' 
 
  WRITE (*, '(A)', ADVANCE='NO') ' Enter argument of Ln[x],         x = ' 
  READ  (*,*) arg 
 
  WRITE (*, '(A)', ADVANCE='NO') ' Enter the number of known terms, N = ' 
  READ  (*,*) N 
 
  IF (N > 137) N = 137   ! ... we like this number ;-) 
 
  x  = arg - one 
  xi = one 
  is = 1 
 
! Initialize S to store the first N+1 known partial sums 
! S0, S1, S2,..., Sn-1, Sn 
! 
! Sn = x + x^2/2 - x^3/3! + x^4/4! - ... + (-1)^n x^n/n! 
! 
  S(0) = zero     ! *** lower bound of the subscript should start at 0 ! *** 
  DO i=1,N 
    xi   = xi*x 
    C(i) = xi/i 
    S(i) = S(i-1) + is * C(i) 
    is   = -is 
  END DO 
 
  WRITE (*, '(A)') ' ' 
  WRITE (*, '(A)') ' ===========================================& 
                    &===========================================' 
  WRITE (*, '(14X,A8,5(X,A12))') 'S(0)', 'S(1)', 'S(2)', 'S(3)',& 
                                 'S(4)', 'S(5)' 
  WRITE (*, '(A)') ' -------------------------------------------& 
                    &-------------------------------------------' 
  WRITE (*, '(A)', ADVANCE='NO') ' before call: ' 
  WRITE (*, '(F8.4,5(X,f12.4))')        S(0:5) 
 
  CALL Limes &   ! call subroutine Limes to calculate Ln[x] 
 (      N,   &   ! in 
   S(0:N),   &   ! inout 
   rLimes,   &   ! out 
   i_Pade,   &   ! out 
   k_Pade,   &   ! out 
      err  )     ! out 
 
! Formated output of results 
! 
  WRITE (*, '(A)', ADVANCE='NO') '  after call: ' 
  WRITE (*, '(F8.4,5(X,f12.4))')        S(0:5) 
  WRITE (*, '(A)') ' ===========================================& 
                    &===========================================' 
  WRITE (*,*) ' ' 
 
  WRITE (*, '(X,A,X,F9.3,X,A,ES12.5)') 'Ln[', arg, '] = ', LOG (arg) 
  WRITE (*, '(X,A,I3,A,I3,A,ES12.5)') 'rLimes[', i_Pade, ',', k_Pade, & 
                                     '] = ', rLimes 
  WRITE (*, '(13X, 1A, ES12.5)')   'err = ', err 
  WRITE (*, '(6X,  1A, ES12.5)') & 
                            'true error = ', ABS ( rLimes - LOG (one + x)) 
 
!----------------------------------------------------------------------------- 
  CONTAINS 
!----------------------------------------------------------------------------- 
!+ Finds the limit of a series 
SUBROUTINE Limes & 
 (      N,       &   ! in 
        S,       &   ! inout 
   rLimes,       &   ! out 
   i_Pade,       &   ! out 
   k_Pade,       &   ! out 
      err  )         ! out 
 
! Description: 
!   Finds the limit of a series in the case where only 
!   the first N+1 terms are known. 
! 
! Method: 
!   The subroutine operates by applying the epsilon-algorithm 
!   to the sequence of partial sums of a seris supplied on input. 
!   For desciption of the algorithm, please see: 
! 
!   [1] T. Mishonov and E. Penev, Int. J. Mod. Phys. B 14, 3831 (2000) 
! 
! Owners: Todor Mishonov & Evgeni Penev 
! 
! History: 
! Version   Date         Comment 
! =======   ====         ======= 
! 1.0       01/04/2000   Original code. T. Mishonov & E. Penev 
! 
! Code Description: 
!   Language:           Fortran 90. 
!   Software Standards: "European Standards for Writing and 
!                        Documenting Exchangeable Fortran 90 Code". 
! 
! Declarations: 
IMPLICIT NONE 
 
!* Subroutine arguments 
!  Scalar arguments with intent(in): 
  INTEGER,   INTENT (IN)    :: N       ! width of the epsilon-table 
 
!  Array arguments with intent(inout): 
  REAL (pr), INTENT (INOUT) :: S(0:)   ! sequential row of the epsilon-table 
 
!  Scalar arguments with intent(out) 
  REAL (pr), INTENT (OUT)   :: rLimes  ! value of the series limes 
  INTEGER,   INTENT (OUT)   :: i_Pade  ! power of the numerator 
  INTEGER,   INTENT (OUT)   :: k_Pade  ! power of the denominator 
  REAL (pr), INTENT (OUT)   :: err     ! empirical error 
 
!* End of Subroutine arguments 
 
!  Local parameters                    ! these two need no description ;-) 
  REAL (pr),    PARAMETER   :: zero = 0.0 
  REAL (pr),    PARAMETER   :: one  = 1.0 
 
!  Local scalars 
  REAL (pr)                 :: A_max   ! maximum element of A 
  INTEGER                   :: i       ! index variable for columns 
  INTEGER                   :: k       ! index variable for rows 
 
!  Local arrays 
  REAL (pr)                 :: A(0:N)  ! auxiliary row of the epsilon-table 
 
!- End of header -------------------------------------------------------------- 
 
!  Parse input: the algorithm cannot employ more elements than supplied on 
!               input, i.e. N <= size(S) 
! 
  IF ( N > SIZE (S(:)) ) THEN 
    WRITE (*, '(A)') '*** Illegal input to Limes: N > size(S)' 
    STOP 1 
  END IF 
 
!  Algorithm not applicable for N < 2 
! 
  IF ( N < 2 ) THEN 
    WRITE (*, '(A)') '*** Illegal input to Limes: N < 2' 
    STOP 2 
  END IF 
 
!----------------------------------------------------------------------------- 
!  I. Initialize with natural assignments 
!----------------------------------------------------------------------------- 
  rLimes = S(N)                    ! the N-th partial sum 
  err    = ABS ( S(N) - S(N-1) )   ! error -> |S(N) - S(N-1)| 
  i_Pade = N                       ! Pade approximant [N/0] 
  k_Pade = 0                       ! 
  A(:)   = zero                    ! auxiliary row initially set to zero 
  A_max  = zero                    ! max. element set to zero 
  k      = 1                       ! algorithm starts from the first row 
 
!----------------------------------------------------------------------------- 
! II. Main loop: fill in the epsilon table, check for convergence ... 
!     (provision against devision by zero employs pseudo-inverse numbers) 
!----------------------------------------------------------------------------- 
  DO 
    IF ( N - 2 * k + 1 < 0 ) EXIT 
 
! Update the auxiliary row A(i) of the epsilon-table 
! by applying the "cross rule". 
! 
    DO i=0, N - 2 * k + 1 
      IF ( S(i+1) /= S(i) ) THEN 
        A(i) = A(i+1) + one/(S(i+1) - S(i)) 
      ELSE 
        A(i) = A(i+1) 
      END IF 
    END DO 
    IF ( N - 2 * k < 0 ) EXIT 
 
!  Update the sequential row S(i) of the epsilon-table 
!  by applying the "cross rule". 
! 
    DO i=0, N - 2 * k 
      IF ( A(i+1) /= A(i) ) THEN 
        S(i) = S(i+1) + one/(A(i+1) - A(i)) 
      ELSE 
        S(i) = S(i+1) 
      END IF 
 
!  Check for convergence, based on A_max; see Ref. [1] 
! 
      IF ( ABS ( A(i) ) > A_max ) THEN 
        A_max  = ABS ( A(i) ) 
        rLimes = S(i) 
        k_Pade = k 
        i_Pade = i + k_Pade 
        err    = one/A_max 
        IF ( S(i+1) == S(i) ) RETURN 
      END IF 
    END DO 
    k = k + 1      ! increment row index 
  END DO 
 
  END SUBROUTINE Limes 
END PROGRAM Test 
\end{verbatim} 
} 

%
\nonumsection{References}

\end{document}